\documentclass[twocolumn,aps,prba,superscriptaddress]{revtex4-1}
\usepackage[utf8]{inputenc}
\usepackage{amsmath}
\usepackage{graphicx}
\usepackage{booktabs}
\usepackage{siunitx}
\usepackage[format=plain]{caption}
\usepackage{subfig}
\usepackage[margin=0.9in]{geometry}
\usepackage{textcomp}
\usepackage{gensymb}
\usepackage{sidecap}
\usepackage[english]{babel}
\usepackage[autostyle]{csquotes}
\usepackage{braket}
\usepackage{array}
\usepackage{color}
\usepackage{multirow}

\begin{document}
\justifying
\captionsetup{justification=Justified,}

\title{Unraveling the Crystallization Kinetics of the Ge$_2$Sb$_2$Te$_5$ Phase Change Compound with a  Machine-Learned Interatomic Potential}

\author{Omar Abou El Kheir}
\affiliation{Department of Materials Science, University of Milano-Bicocca, Via R. Cozzi 55, I-20125 Milano, Italy}
\author{Luigi Bonati}
\affiliation{Atomistic Simulations, Italian Institute of Technology, Via Enrico Melen 83, 16152, Genova, Italy}
\author{Michele Parrinello}
\affiliation{Atomistic Simulations, Italian Institute of Technology, Via Enrico Melen 83, 16152, Genova, Italy}
\author{Marco Bernasconi}
\affiliation{Department of Materials Science, University of Milano-Bicocca, Via R. Cozzi 55, I-20125 Milano, Italy}

\begin{abstract}
The phase change compound Ge$_2$Sb$_2$Te$_5$ (GST225) is exploited in advanced non-volatile electronic memories and in neuromorphic devices which both rely on a fast and reversible transition between the crystalline and amorphous phases induced by Joule heating. The crystallization kinetics of GST225 is a key functional feature for the operation of these devices. We report here on the development of a machine-learned interatomic potential for GST225 that allowed us to perform large scale molecular dynamics simulations (over 10000 atoms for over 100 ns) to uncover the details of the crystallization kinetics in a wide range of temperatures of interest for the programming of the devices. The potential is obtained by fitting with a deep neural network (NN) scheme a large quantum-mechanical database generated within Density Functional Theory. The availability of a highly efficient and yet highly accurate NN potential opens the possibility to simulate phase change materials at the length and time scales of the real devices. \\

 \textbf{This version of the article has been accepted for publication, after peer review but is not the Version of Record and does not reflect post-acceptance improvements, or any corrections. The Version of Record is available online at: https://doi.org/10.1038/s41524-024-01217-6}
 
\end{abstract}




\maketitle

\section{INTRODUCTION} 
%\label{sec1}
\vskip 0.5 truecm 
In the last decades, the  rise of the demand for  data processing and storage has stimulated a strong effort in the search of new computing architectures and memory devices. Chalcogenide phase change materials \cite{wuttig2007phase} are at the heart of some of the most mature technologies suitable to respond to these needs.
Indeed, these materials are exploited in both emerging non-volatile electronic memories, named  phase change memories (PCM) \cite{wuttig2007phase,noe2017phase,fantini}, and in neuromorphic and in-memory computing devices \cite{kuzum2012nanoelectronic,tuma2016stochastic,sebastianNanotech}.
PCMs rely on a rapid (down to tens of ns) and reversible transformation induced by Joule heating between the crystalline and amorphous phases of the prototypical phase change compound Ge$_2$Sb$_2$Te$_5$ (GST225) \cite{wuttig2007phase}.  Read out of the memory consists of the measurement of the resistance of GST225 which differs by about three orders of magnitude in the two phases \cite{wuttig2007phase}. Ge-rich GeSbTe alloys with  crystallization temperatures much higher than that of GST225 have also been investigated for memories embedded in microcontrollers for automotive applications \cite{Cappelletti_2020,ZULIANI201527}. Moreover, partial crystallization of the amorphous phase leads to different levels of resistivity which is exploited in the realization of artificial synapses for neuromorphic and in-memory computing \cite{sebastianNanotech,zhang2019designing}.
\\
 A key functional property for all these applications is the crystallization kinetics of the amorphous phase between the glass transition (T$_g$) and the melting (T$_m$) temperatures. This feature is, however, difficult to be investigated experimentally because of the very high nucleation rates and crystal growth velocities (a few m/s).  Indeed, ultrafast differential scanning calorimetry (DSC) was needed to measure the crystal growth velocity at the high temperatures of interest for the operation of the devices \cite{orava}.   Information on the crystallization kinetics was, however, inferred from DSC under several assumptions on the crystallization mechanisms based on classical nucleation theory  which require further validations.
 \\
On the other hand, atomistic insights on the early stage of the crystallization process have been  provided by molecular dynamics (MD) simulations based on density functional theory (DFT) \cite{zhang2019designing,elliot,PhysRevLett.107.145702,kalikkacrys3,ronneberger2015crystallization,ronneberger2018crystal,GST124,rao2017reducing,elliott2017}.
Several works on GST225  revealed that the high nucleation rate can be ascribed to the presence in the supercooled liquid of four-membered rings, which  are the same building blocks of the cubic rocksalt crystal, that act as seeds for the formation of crystalline nuclei \cite{zhang2019designing,elliot}. Moreover, the high crystal growth velocity is ascribed to the high fragility of the supercooled liquid which can sustain high atomic mobility down to temperatures close to T$_g$, where the  thermodynamical driving force for  crystallization is also high \cite{zhang2019designing}. Although they provided  crucial information on the early stage of crystallization, DFT-MD  methods suffer from limitations in system size and in simulation time that prevent to address some important issues for the operation of the memory devices.  A DFT-MD  study of the crystallization kinetics on an extended temperature range to test the applicability of classical nucleation, for instance,  is  still lacking for GST225. 
\\
In the last few years, the development of interatomic potentials based on the fitting of a large DFT database by machine learning techniques emerged as a viable approach to  overcome these limitations of DFT-MD and enlarge  the scope of  DFT methods \cite{Behler,behler2016perspective,deringer2019machine,wang2018deepmd}. 
Concerning  phase change materials, machine learning schemes based on  Neural Network (NN) methods  have been exploited  to study the crystallization of the phase change materials GeTe \cite{SossoNN,sosso2013,LEE2020109725} and Sb \cite{dragoni2021mechanism,SHI2021106146}.  NN simulations of GeTe also allowed addressing the study of the aging of the amorphous phase \cite{Gabardi2015} that leads to an increase of the electrical resistance with time (drift) which is detrimental for the operation of the memory devices \cite{Raty2015}. 
For GST225, a machine learning interatomic potential was recently developed with the Gaussian  Approximations Potential (GAP) method \cite{mocanu2018modeling} which, however, suffers from some inaccuracies in reproducing the DFT results on the structural properties of the amorphous phase such as the fraction of homopolar bonds which are believed to play a crucial role in the aging process \cite{Gabardi2015,Raty2015}.
 Improvements of this potential have been, however, very recently achieved \cite{DeringerNatureEl}.
\\
In this paper, we report on the development of an interatomic potential for GST225 within the NN scheme implemented in the DeePMD code \cite{wang2018deepmd,PhysRevLett.120.143001}. We first validated the potential  on the structural, dynamical and thermodynamical properties of the liquid, amorphous and crystalline phases. Then, we employed the NN potential to study the crystallization kinetics over the wide temperature range of interest for the operation of the devices, aiming at assessing the applicability of classical nucleation theory. 
 The simulations reveal that crystallization kinetics in the temperature range 500-650 K is diffusion-controlled with an activation energy corresponding to that of the self-diffusion coefficient. We also show that a modified form \cite{jackson} of the phenomenological Wilson-Frenkel formula \cite{WF1,FrenkelJ} is suitable to fit the data on a wider temperature range.

We remark that the efficient implementation of the DeePMD scheme allows simulating tens of thousands of atoms for tens of nanoseconds at an affordable computational cost. We envisage that this feature could be further exploited to address some important issues for the operation of ultrascaled devices such as the effect of confinement and nanostructuring on the crystallization kinetics \cite{KooiReview}, or  the possible existence of a strong-to-fragile transition in the supercooled liquid close to T$_g$ which is particularly relevant for the aging of the amorphous phase \cite{PriesStrongFragile}, just to name a few.

\section{RESULTS}
\subsection{Generation and Validation of the Neural Network Potential}

The NN potential was obtained by fitting DFT (see Methods) energies, forces and the stress tensor of a database containing about 180000 supercell models (configurations) of GST225 in the liquid, amorphous, cubic and hexagonal phases by  using the DeePMD-kit open-source package \cite{wang2018deepmd,PhysRevLett.120.143001}. 
\\
Initially, the training set consisted of  about 5000 configurations. Then, the NN potential was refined by expanding the database in an iterative process with atomic configurations generated from DFT-MD trajectories to enhance the description of specific properties and from NN-MD trajectories whose energy was badly described by the intermediate versions of the potential. In the final database of about 180000 configurations, we covered a wide range of thermodynamical conditions, i.e., different temperatures up to 2000 K and several densities in the range 0.026-0.036 atom/\AA${^3}$. 
Details  of configurations in the database at different conditions are reported in Tab. \ref{tab:configData} while information on the NN scheme are given in section on Methods.  Each configuration refers to the DFT energy, forces and stress of  108-atom cells  for the liquid, amorphous and hexagonal crystalline phases and of a 57- or 98-atom cell for the cubic crystal. The configurations of the supercooled liquid  were extracted from simulations of several tens of ps at fixed temperature in the range 500-900 K, {\sl  quenching}  refers to simulations in which the system was cooled very rapidly  from 990 to 300 K and {\sl  metadynamics}  refers to biased simulations \cite{laio} to enhance the sampling in a specific region of the phase space.
The metadynamics method  is based on the identification of appropriate order parameters or collective variables (CVs) that describe the slow modes of the process. An external potential is then added to enhance the fluctuations of the selected CVs. The method allows sampling the free energy surface by overcoming activation barriers much larger than the thermal energy in a short time span. As a result, it allows to collect a more heterogeneous set of configurations than standard MD simulations, obtaining robust potentials that can also describe phase transitions~\cite{bonati2018silicon,niu2020ab}. In particular, we performed metadynamics simulations in the supercooled liquid by using the  coordination numbers as CVs.
The addition of such configurations to the training set was  crucial to reproduce the fraction of homopolar bonds in the amorphous phase.\\
%\newpage
\begin{table}[t]
\begin{center}
\captionsetup{type=Table}
\captionof{table}{Details of the database used for the training of the NN potential.}
\label{tab:configData}
\begin{tabular}{c c}
\hline
phase/simulations & number of configurations \\
\hline
liquid    & 4489\\
supercooled liquid  &64440\\
quenching  		 	&70463\\	 			
amorphous   &3620\\
cubic crystal    &8764 \\ 
hexagonal crystal  &7935\\
metadynamics  &18524\\
\hline
\\
\end{tabular}
\end{center}{}
\end{table}

\begin{figure*}
 \centering
 \includegraphics[width=\textwidth, keepaspectratio]{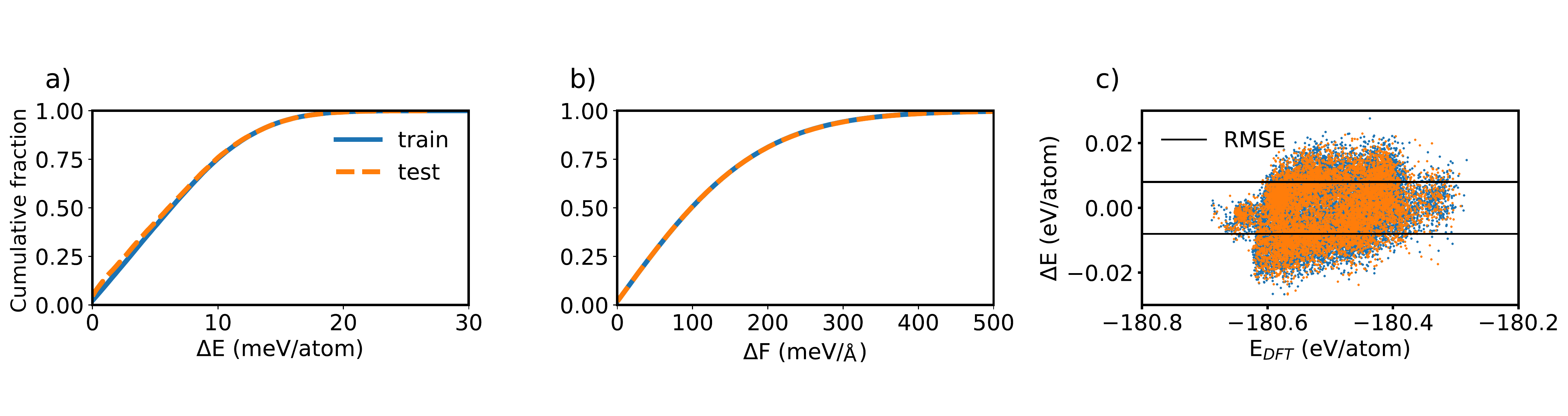}
\caption{ \textbf{Accuracy of the neural network potential.} Cumulative fraction of the absolute errors of the NN potential in training and test data sets for a) the energies per atom ($\Delta $E=$|$E$_{DFT}$-E$_{NN}$$|$) and b) forces ($\Delta $F=$|$F$_{DFT}$-F$_{NN}$$|$). c)  The distribution of the absolute error as a function of DFT energy.}
 \label{fig:erroriNN} 
\end{figure*}

The energy root-mean-square error (RMSE) between the DFT values and those predicted by the NN potential is 8.4 and 8.6 meV/atom for the training and test sets. The RMSE on forces is 159 meV/\AA\ for both sets. The distributions of NN errors on energies and forces are given in Fig. \ref{fig:erroriNN}. The typical average  error obtained with DeePMD for highly disordered phases of multicomponent systems like ours (i.e  liquid and/or amorphous phases)  are in the range 2-7 meV/atom and 90-145 meV/\AA\ \cite{RMSE1,RMSE2,RMSE3,RMSE4}. The NN potential has been validated on the properties of the liquid, amorphous, cubic and hexagonal phases as described in the separate sections below. 
All the NN-MD simulations were performed with the LAMMPS code \cite{LAMMPS} and a Nos\'e-Hoover thermostat \cite{noseart,hoover}.  
\\

{\sl The Liquid Phase:}
The structural properties of the liquid phase of GST225 have been obtained from NN-MD simulation with a 999-atom supercell at 990 K and compared with those obtained from DFT-MD simulation with a 216-atom supercell at the same temperature. In both models, we used the experimental density of the amorphous phase of 0.0309 atoms/\AA$^3$ \cite{njoroge2002density} which is very close to that of the liquid at 893 K (0.0307 atoms/\AA$^3$ \cite{mazzarelloliq}). To generate the liquid models we first performed a 7 ps long MD simulation at 2000 K to properly randomize the atoms in the box. Then, we performed a second equilibration at 990 K for 10 ps. Finally, we evaluated the structural properties over trajectories 10 ps long. 

\begin{figure*}
 \centering
 \includegraphics[width=\textwidth, keepaspectratio]{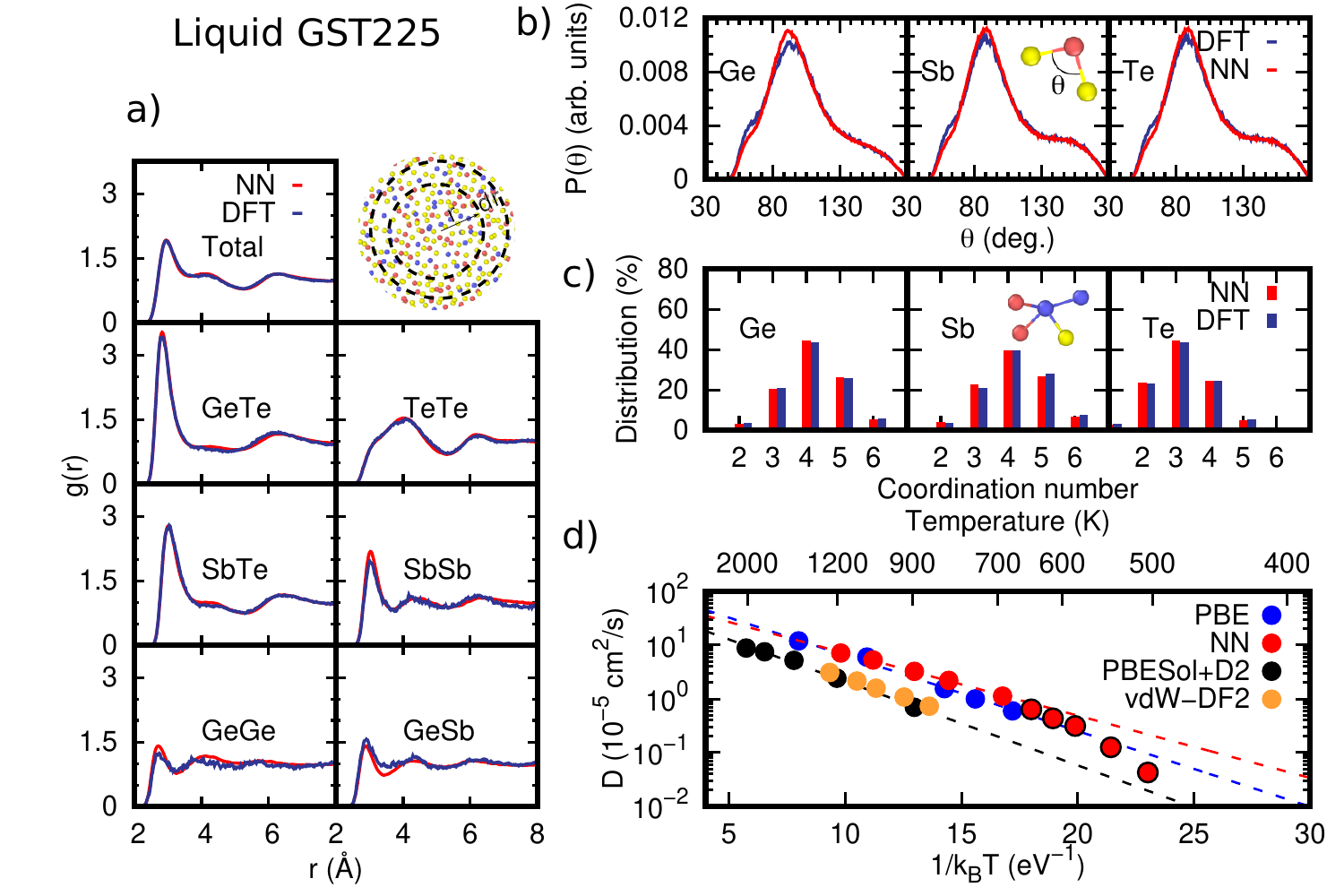}
 \caption{ \textbf{Structural  and dynamical properties of the liquid phase.}
 Structural  and dynamical properties of GST225 in the liquid phase from  DFT (blue curves) and NN (red curves) simulations with a 216-atom or 999-atom models, respectively.
 a) Total and partial radial distribution functions at 990 K. The position of the first maximum and minimum of each partial correlation function are given in the Supplementary Table 1. b) Angle distribution function resolved per central atomic species at 990 K. The data were normalized to the number of triplets in each model. c) Distribution of coordination numbers resolved per chemical species at 990 K computed by integrating the partial pair correlation functions up to a bonding cutoff which corresponds to 3.2 \AA\ for all pairs except for Sb-Te for which we use 3.4 \AA\ as was done in Ref. \cite{caravati2009first}. d) The
Arrhenius plot of the diffusion coefficient $D$ as a function of temperature compared with previous DFT results with the different functionals PBE \cite{caravatiDIFF}, PBESol \cite{MicoulautDIFF} and vdW-DF2 \cite{mazzarelloliq}) (see Supplementary Table 2 for the corresponding Arrhenius parameters). The NN data of $D$ below 700 K are shown with a black border as they have been excluded from the Arrhenius fit.}
 \label{fig:lgdr}
\end{figure*}

The  pair correlation functions,  the bond angles distribution function and the distribution of coordination numbers are compared with DFT results in Fig. \ref{fig:lgdr}.   The structural properties obtained from NN-MD simulations are in excellent agreement with those obtained from DFT-MD which suggests that the NN potential reliably describes the liquid phase.

Regarding the dynamical properties, we computed the self-diffusion coefficient $D$ from NN-MD simulations at several temperatures above T$_m$  and below T$_m$ in the supercooled liquid spanning the range 500-1200 K. The density was fixed to the value of the experimental amorphous phase (0.0309 atoms/\AA$^3$) as it was done in the previous DFT-PBE work \cite{caravatiDIFF} that we take as a reference for the validation. The self-diffusion coefficients was obtained from the mean square displacement (MSD) and the Einstein relation MSD=6$D$t from equilibrated trajectories at constant energy over time intervals from 40 ps at high temperatures to 300 ps at low temperatures. At temperatures above 700 K the data can be fitted 
 with the Arrhenius function $D$=$D_0$exp$({-E_a}/{k_BT})$ (see Fig. \ref{fig:lgdr}d) with  E$_a$= 0.267 eV and $D_0$=1.03x10$^{-3}$ cm$^2$/s which are similar to other DFT values reported in the literature (see  Supplementary Table 2). 
 Below 700 K deviations from the Arrhenius law are present due to the fragility of the system.  For a fragile liquid, 
  the self-diffusion coefficient  can  be fitted  in a wider range of temperatures with the    Cohen-Grest (CG) formula \cite{CG}  as $\log_{10}(D(T))=A-{2B}/(T-T_0 + [(T-T_0)^2 + 4CT ]^{1/2})$, which for GST225 yields 
  A=-2.45, B=602 K, C=17.3 K and $T_0$=330.6 K (see  Supplementary Figure 1).  We have chosen the CG formula because it was used in the experimental work in Ref. \cite{orava}  to fit 
the kinetic prefactor in the crystal growth velocity inferred from DSC data  on which we will come back later in the discussion of the crystallization kinetics. The diffusion coefficient as a function of temperature resolved for the different atomic species is also reported in the Supplementary Figure 1.

\begin{figure*}
 \centering
 \includegraphics[width=\textwidth, keepaspectratio]{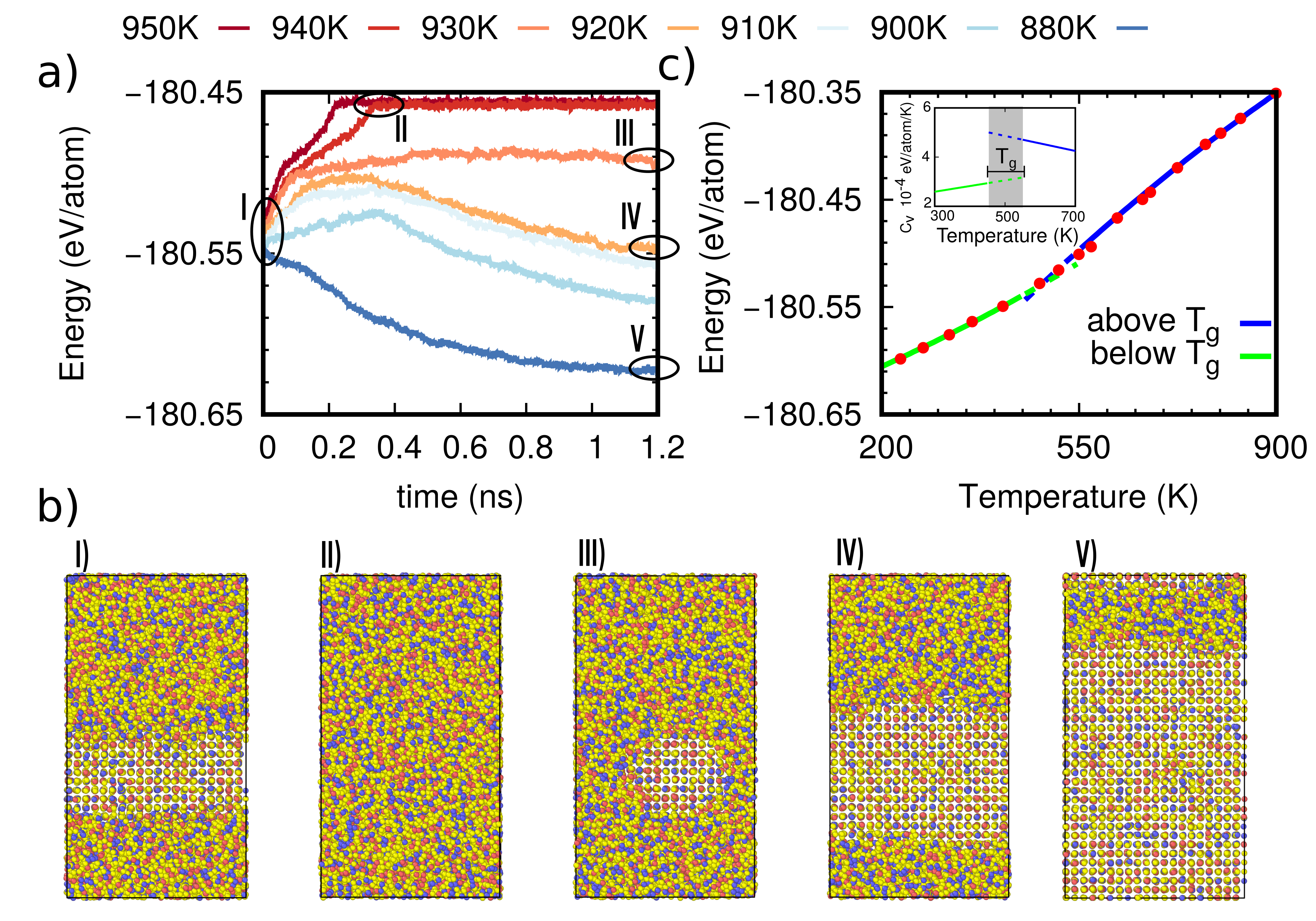}
 \caption{\textbf{ Melting and glass transition temperatures.} a)  The potential energy as a function of time at different temperatures of the cubic-liquid interface model.
 Simulations are performed at constant volume at a density of 0.0309 atom/\AA$^3$ which corresponds to the theoretical equilibrium density of the cubic crystal (see Sec. 2.3). The simulation cell with edges  6.15$\times$6.15$\times$11.08 nm$^3$ was initially prepared with an interface lying on the $xy$ plane separating a 2 nm thick slab in the cubic phase from a  liquid slab 9 nm thick.
 b) Snapshots along the trajectories of panel a) showing the initial configuration and the movement of the interface between the two phases. The melting temperature is estimated to be in the range 920-930 K.
 c) Energy (potential plus kinetic) of the supercooled liquid as a function of temperature in simulations at constant volume.  The lines are  a quadratic fit below and above T$_g$. The resulting heat capacity  at constant volume is given in the inset. The data at each temperature are obtained by averaging   over 40 ps simulations of a 999-atom cell,  initially equilibrated at 1000 K and then cooled  down to the target temperature in 20-100 ps. }
 \label{fig:diff}
\end{figure*}

As a further step in the validation of the NN potential, we estimated the melting temperature of the crystalline cubic phase of GST225 by means of the phase coexistence method \cite{allen2017computer}. To this aim, we prepared a 12960-atom model of the cubic-liquid interface initially set at 900 K and at the theoretical equilibrium density of the cubic phase.  
The crystal-liquid interface corresponds to the (001) surface of the cubic crystal.
Then we carried out several independent simulations at constant volume and at different temperatures in the range 880-950 K to monitor the potential energy as a function of time (Fig. \ref{fig:diff}a), starting from the initial configuration in Fig. \ref{fig:diff}b. At temperatures above T$_m$ we expect the crystalline region to melt as it is the case at  950 and 940 K (see snapshot II in Fig. \ref{fig:diff}b ) while for temperatures below T$_m$ we expect  the crystalline region to grow, as it is indeed the case  below 920 K. This set of simulations suggests that the melting point of the NN potential is within the range 920-930 K which is very close the experimental value of 900 K \cite{mazzarelloliq}. As stated above, these results refer to simulations with the density fixed to that of the cubic phase, also for the liquid.  A better estimation of T$_m$ should, however, be obtained from NPT simulations to describe the density change across melting and the thermal expansion of the crystal. NPT simulations with the PBE functional are, however, problematic as they underestimate significantly the equilibrium density due to the coalescence of nanovoids
as it was observed for GeTe \cite{sosso2012breakdown}. 
We then repeated the same calculations in the temperature range  925-940 K by adding van der Waals interactions (vdW) according to Grimme \cite{D2} (D2) which prevents the coalescence of nanovoids that form by decreasing the density as it was also reported for GeTe \cite{sosso2012breakdown}. Starting from the same initial configuration, we first carried NPT-MD simulation allowing the cell edges to change at fixed angles. The equilibrium density is reached on a time scale of 30 ps  which is much shorter than that required for crystallization. From several NPT simulations at different temperature we estimated T$_m$=940 K which is close to the previous NVT results with no vdW interactions (see  Supplementary Figure 2). At this temperature, the latent heat of melting is $\Delta H_m$=163 meV/atom which is close to the experimental values of 120 or 173 meV/atom \cite{melting1,melting2}. To calculate the latent heat at 940 K, we performed NPT simulation for a 999-atom model of the liquid and a 900-atom  special quasi-random-structure model \cite{sqs} for the cubic crystalline phase.\\

To assess the error in the latent heat introduced in case one does not take into account the density change upon melting (i.e. NVT simulations), we also computed the energy of the liquid phase at the equilibrium density of the cubic crystalline phase at 940 K. The resulting heat of melting at constant volume is $\Delta E_m$ = 154     meV/atom. The value of $\Delta E_m$ computed at constant volume without vdW interactions is instead 166 meV/atom. We expect this latter value to differ from the latent heat computed in the NPT ensemble by a similar error of about 10 meV/atom as found for the simulations with vdW interactions.
We also estimated the change in T$_m$ due to the choice of NVT conditions by using the Clausius-Clapeyron equation  with the  calculated latent heat at 940 K, and the theoretical equilibrium density of the two phases within NN+D2 simulations. This yields a change  in T$_m$ of about 10 K.

All the results reported in the following still refer to simulations without vdW interactions as  we want to validate the NN potential over DFT-PBE data. vdW interactions might be later added in simulations with the NN potential even by using different schemes.

As a final result on the supercooled liquid, we report in Fig. \ref{fig:diff}c  the energy as a function of temperature computed  at a constant density of 0.0309 atoms/\AA$^3$. The energy displays clearly two slopes as a function of temperature which give rise to a jump in the specific heat shown in the inset of Fig. \ref{fig:diff}c. The jump in C$_v$ at about 500 K can be identified with the glass transition temperature which turns out to be very similar to the latest experimental value of T$_g$=473 K  reported in Ref. \cite{pries2019switching},  where it was also proposed that crystallization in DSC occurs below T$_g$.
The decrease of C$_v$ with temperature in the supercooled liquid above T$_g$ is another  feature typical of fragile liquids \cite{angell1995formation}.
We also repeated the same simulations at constant (zero) pressure with the NN+D2 potential, the resulting enthalpy and $C_p$ as a function of temperature, reported  in the Supplementary Figure 3, yield a very  similar estimate of T$_g$.

\begin{figure*}
 \centering
 \includegraphics[width=\textwidth, keepaspectratio]{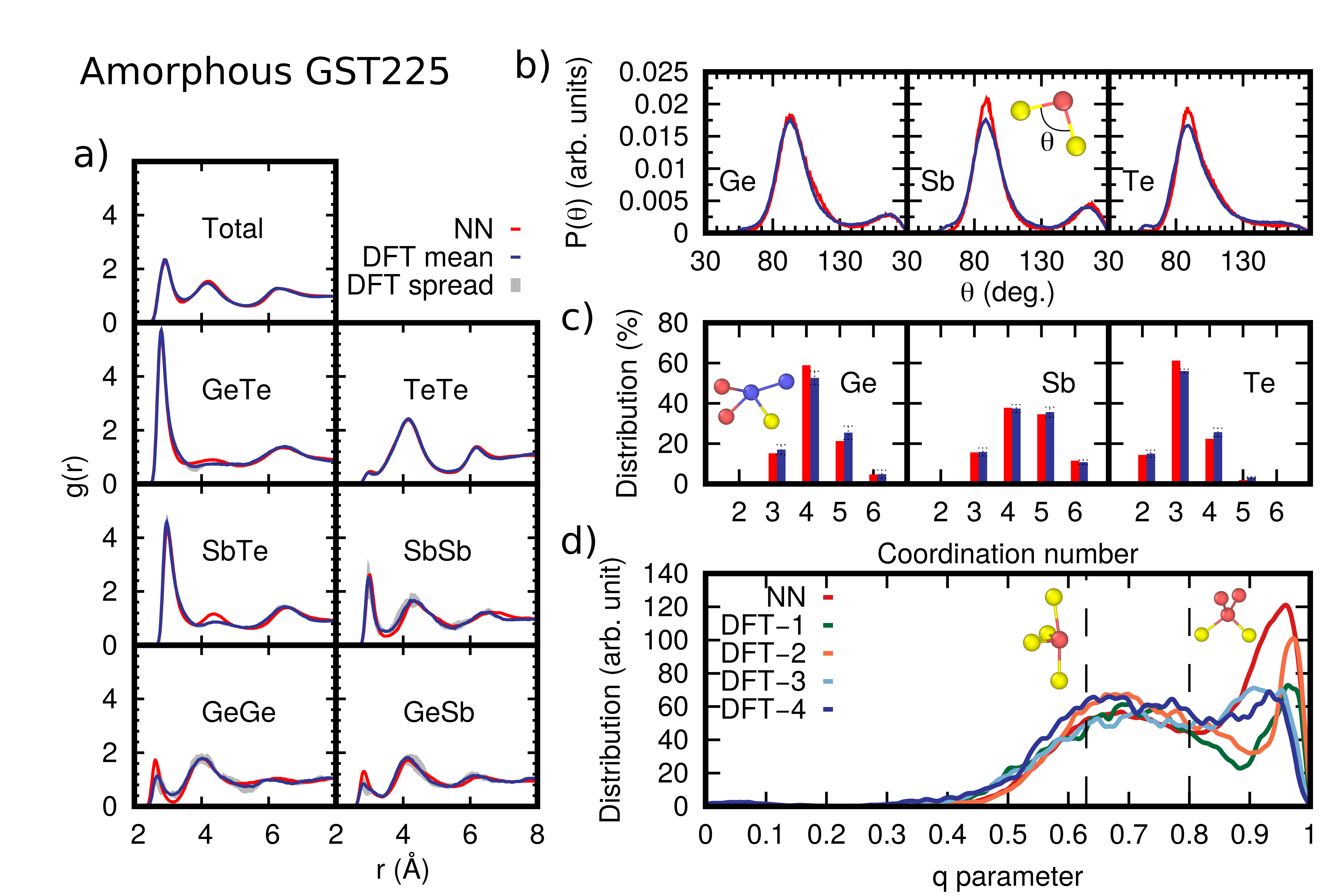}

 \caption{ \textbf{Structural properties of the amorphous phase.} Structural properties of amorphous GST225 at 300 K from NN (red curves) and DFT (blue curves) simulations. 
 DFT data are averaged over four independent 216-atom models while NN data are obtained from a 999-atom model. a) Total and partial radial distribution functions. The position of the first maximum and minimum of the different functions is reported  in the Supplementary Table 3.  b) Angular distribution function resolved per central atomic species. The data are normalized to the number of triplets in each model. c) Distribution of coordination numbers resolved per chemical species, obtained by integrating the pair correlation functions up to a bonding cutoff corresponding to 3.2 \AA\ for all pairs except for Sb-Te for which a longer cutoff value of 3.4 \AA\ was used. The spread over the four DFT models are indicated by error bars.  
 d) Distribution of the local order parameter $q$ for tetrahedricity for four-coordinated Ge atoms for the NN model  and for each of our four DFT models.}
 \label{fig:agdr}
\end{figure*}

{\sl The Amorphous Phase:}
The NN potential was then validated on the structural properties of the amorphous phase. A 999-atom model of amorphous GST225 at the experimental density (0.0309 atom/\AA$^3$) was generated by quenching from the melt at 990 K to 300 K in 100 ps. 
The  pair correlation functions, the bond angles distribution functions and the distribution of the coordination numbers are compared in Fig.\ref{fig:agdr} with DFT results from four 216-atom models 
 generated with the same quenching protocol and at the same density of the NN model.
 The resulting average partial coordination numbers are compared in Table 2. The comparison of the structural properties of two NN models  of different sizes containing 999 or 7992 atoms (see  Supplementary Figure 4)  shows that the structural properties are well converged in the 999-atom cell.
 The agreement between NN and DFT simulations is overall excellent, including small but very important  details such as the fraction of homopolar Ge-Ge and Sb-Sb bonds.

\begin{table*}[t]

\captionsetup{type=Table}
\captionof{table}{Average coordination numbers for different pairs of atoms in amorphous GST225 at 300 K generated from  DFT (in parenthesis) and NN simulations.
Error bars are obtained  from the analysis of four models.}
\label{tab:am225NN}

\begin{tabular}{ c c c c c c  }
\hline

-       & Ge                                                             &                                                            Sb & Te \\ \hline
Total   & $4.19\pm$0.01 (4.17$\pm$0.1) &4.46$\pm$0.02 (4.40$\pm$0.05)&3.16$\pm$0.02 (3.16$\pm$0.05)\\
With Ge &0.33$\pm$0.04 (0.34$\pm$0.05)&0.33$\pm$0.02 (0.26$\pm$0.05)&1.41$\pm$0.02 (1.42$\pm$0.02)\\
With Sb & 0.33$\pm$0.02(0.26$\pm$0.05) &0.51$\pm$0.03 (0.51$\pm$0.1) &1.45$\pm$0.02 (1.45$\pm$0.05)\\
With Te & 3.53$\pm$0.06 (3.57$\pm$0.05)&3.64$\pm$0.05 (3.63$\pm$0.1) &0.29$\pm$0.02 (0.28$\pm$0.02)\\
\hline
\\
\end{tabular}
\end{table*}

\begin{figure*}

 \centering
\includegraphics[width=\textwidth, keepaspectratio]{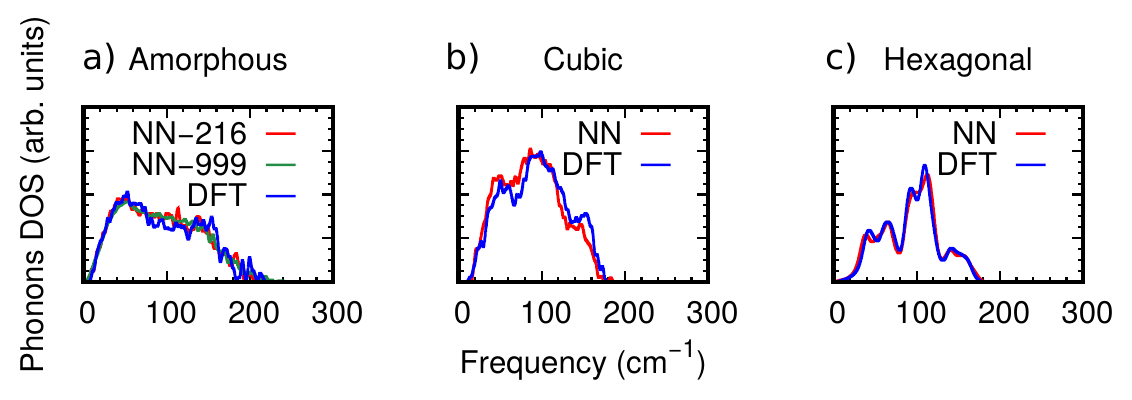}
 \caption{ \textbf{Phonons.}
 a) Phonon density of states (DOS) of the amorphous phase obtained from $\Gamma$-point frequencies of a 216- and 999-atom supercell with the NN potential and of a 216-atom supercell with DFT.
  b) DFT and NN phonon DOS of the cubic phase of GST225. The phonons were computed at the $\Gamma$- point of a 270-atom supercell.    Phonon frequencies in panel a) and b) are obtained by diagonalizing the dynamical matrix  from finite atomic displacements. Each frequency is then broadened by a Gaussian function with a width of 1 cm$^{-1}$.  c) Phonon density of states of the crystalline hexagonal phase of GST225 computed with the NN+D2 potential and by DFT+D2 from Ref. \cite{campi} (Kooi phase, see text). The NN+D2 DOS for the hexagonal phase are obtained from the the force constant matrix of a 12x12x4 supercell  and the Phononpy code \cite{phonopy}. }
 \label{fig:a225qparam}
\end{figure*}

The structure of amorphous GST225 is similar to that emerged in previous DFT works \cite{caravati2007coexistence,akola2007structural}.
Te atoms are mostly three-fold
coordinated in a pyramidal geometry, Sb atoms are both three-fold coordinated in a pyramidal geometry (three bonding angles of 90$^{\circ}$) and four- or five-fold
coordinated in a defective octahedral environment (octahedral
bonding angles but coordination lower than six), most of the
Ge atoms are in pyramidal or defective octahedral geometry
while a minority fraction of Ge atoms are in tetrahedral geometries. 
The bonding geometry is revealed by the coordination numbers and by the angle distribution function where the peak at about 90$^o$  is due to pyramidal and defective octahedral configurations, while the weak
peak at about 170$^o$ is due to axial bonds in  defective octahedra. The shoulder at about 109$^o$ is due to tetrahedra which are favored by homopolar Ge-Ge bonds \cite{deringer2014bonding} that are present in the liquid and survive in the amorphous phase due to fast quenching.
A quantitative measure of the fraction of tetrahedral environments can be obtained from the local order parameter $q$ introduced in 
Ref.~\cite{qparam}. It is defined as $q=1-\frac{3}{8}\sum_{i > k} (\frac{1}{3} + \cos \theta_{ijk})^2$,   where the sum runs over the pairs of atoms bonded to a central atom $j$ and 
forming a bonding angle $\theta_{ijk}$.The order parameter evaluates to $q$=1 for the ideal tetrahedral geometry, to $q$=0 for the 6-fold coordinated octahedral site,
to $q$=5/8 for a 4-fold coordinated defective octahedral site, and $q$=7/8 for a pyramidal geometry. 
The distribution of the local order parameter $q$ for tetrahedricity  is reported in Fig. \ref{fig:agdr}d for four-coordinated Ge atoms. The bimodal shape corresponds to tetrahedral and defective octahedral geometries. We quantified the fraction of Ge atoms in a tetrahedral environment by integrating the $q$ parameter between 0.8 and 1 as discussed in  previous works \cite{spreafico}. In the NN model, 30\% of Ge atoms are in the tetrahedral geometry to be compared with an average value of 23\%  in our DFT models. DFT calculations in literature report a fraction of tetrahedral Ge in the range 27-35\% \cite{caravati2007coexistence,akola2007structural}. 
We remark, however, that a very recent paper \cite{Massobrio2023} reports on the comparison of two 504-atom models of amorphous GST225 generated with two different pseudopotentials for Ge, based either on the Troullier-Martins (TM)\cite{Troullier} or GTH schemes. Although the two models reproduce similarly well the experimental neutron diffraction data, the fraction of tetrahedral Ge is 36 $\%$ in the GTH model while it goes up to 65 $\%$ in the TM model which better reproduces the Ge-Ge and Ge-Te bond lengths inferred from extended-x-ray-fine-structure measurements \cite{Massobrio2023}. A similarly larger fraction of tetrahedra have been obtained in Ref. \cite{Micoulaut2021}.
Information on the medium range order is provided by the distribution of rings length which in GST225 is dominated by the four-membered rings \cite{elliot,caravati2007coexistence}. This feature has been regarded as the precursor for rapid crystallization as the  four-membered ring is also the structural unit of the rocksalt crystalline phase \cite{elliot}.
The distribution of rings length reported in the Supplementary Figure 5 for the NN and DFT models shows that the 
NN potential is able to reproduce also this crucial feature. 
Another useful descriptor of the structure of the amorphous phase is given by the angular-limited three-body correlation function (ALTBC) \cite{Bichara} which highlights the presence of a short and a long axial bond in the defective octahedral configurations \cite{mazzarelloliq}. The ALTBC functions for NN and DFT simulations of amorphous GST225 are in good agreement as well (see  Supplementary Figure 6). 
The NN potential is thus able to reproduce very well the structural properties of the amorphous phase including very crucial details such as the fraction of Ge atoms in tetrahedral configurations and the fraction of homopolar Ge-Ge bonds which might rule the aging of the amorphous phase and the consequent increase (drift) of the electrical resistance with time, as it occurs in the amorphous phase of the parent compound GeTe \cite{Gabardi2015,Raty2015}. Concerning the dynamical properties, the NN  potential reproduces well the DFT phonon density of states (DOS) of amorphous GST225 as shown in  Fig. \ref{fig:a225qparam}a.

{\sl The Crystalline Phases:}
At normal conditions, GST225 crystallizes in a hexagonal phase (space group $P{\bar 3}m1$)\cite{petrov,kooi,matsunaga}. 
However, the amorphous phase crystallizes in a metastable cubic rocksalt crystal which is the phase of interest for the operation of the memory devices. The
metastable cubic phase  consists of a NaCl structure with the anionic sublattice occupied by Te and the cationic one occupied by Ge, Sb and 20\% of vacancies.
In this benchmark, the cubic phase was modeled by  a 300-site supercell (the same as in Ref. \cite{caravati2009first}) with 30 vacancies and 270 atoms at the stoichiometric GST225 composition. 
The hexagonal phase  contains instead 9 atoms  in the primitive unit cell arranged along the $c$ direction with a  ABCABC stacking.  Each formula unit forms a lamella separated from the others by a so-called vdW gap, although the interlamella interaction is not just a vdW contact  as discussed in Ref. \cite{wuttig2018incipient}. 
Three different models of hexagonal GST225 have been proposed in literature differing in the distribution of Sb/Ge atoms in the cation sublattices \cite{petrov,kooi,matsunaga}. Here, we considered the Kooi stacking \cite{kooi} with Sb atoms occupying the cation planes close to the vdW gap. 
We computed the equation of state at zero temperature of the cubic and 
hexagonal phase by fitting the energy-volume points by the Birch-Murnaghan formula (see  Supplementary Figures 7-8). The resulting parameters at equilibrium  are compared  with DFT  data in Tab. \ref{tab:pEOScubic}. 
  The theoretical NN (and DFT-PBE) equilibrium density of the cubic phase (0.0309 atom/\AA$^3$) turns out to be equal to the experimental density of the amorphous phase. By adding vdW interactions (D2)\cite{D2} the equilibrium density of the cubic phase raises to 0.0332 atom/\AA$^3$ which is closer to the experimental value of 0.0328 atom/\AA$^3$ \cite{matsunaga}.
 The NN potential reproduces well also the phonon DOS
 obtained by DFT as shown 
in Fig. \ref{fig:a225qparam}b,c. In the hexagonal phase,  GST225 features phonon  instabilities when the DFT-PBE scheme is applied \cite{campi} which, as expected, are also present in NN calculations as it should be. These instabilities at the PBE level are removed by including the semiempirical vdW correction due to Grimme (D2) \cite{campi}. Hence,
we calculated the phonon dispersion relations with the NN+D2 potential by employing the Phononpy code \cite{phonopy}. The resulting phonon  DOS
  are compared with previous DFT+D2 results \cite{campi} in Fig. \ref{fig:a225qparam}c, while 
 the phonon dispersion relations 
are compared with DFT+D2 results \cite{campi}  in the Supplementary Figure 9.
Although reasonable, the agreement with DFT results is less satisfactorily for the phonon dispersion relations than for all the other properties analyzed so far.  In fact, it is usually rather difficult to reproduce phonon dispersion relations by a NN potential not explicitly devised for this property.

In summary, although the RMSE for the energies and forces are not very small (8 meV/atom and 159 meV/\AA), albeit similar to other NN potentials in literature for disordered multicomponent materials, the validation of the potential over the properties of  liquid, amorphous and crystalline phases is excellent.  Overall, we judge that our potential is sufficiently accurate to address the study of many properties of this system including the crystallization kinetics which is the subject of the next section.

\begin{center}
\captionsetup{type=Table}
\captionof{table}{Fitting parameters of the Birch-Murnaghan equation of state of cubic and hexagonal GST225  from NN and DFT-PBE calculations at zero temperature. The energy (E$_0$), volume (V$_0$), bulk modulus (B) and derivative of B (B$^\prime$) at equilibrium are reported.}
\label{tab:pEOScubic}
\resizebox{\columnwidth}{!}{\begin{tabular}{c c c c c}

\hline
 & E$_0$ (eV/atom)& V$_0$ (\AA$^3$\//atom)& B (GPa) & B$^\prime$ \\
\hline\\
DFT cubic & -180.7424 & 32.42&25.1&7.53\\
NN cubic & -180.7408 & 32.35&22.5&9.41\\
DFT hexagonal & -180.7883 & 31.13 &19.8&21.75\\
NN hexagonal & -180.7890 & 32.13 &18.7&23.38\\
\hline

\\
\end{tabular}}

\end{center}{}

\subsection{Simulation of the Crystallization Process}
As a first application of the NN potential for GST225, we studied the kinetics of the crystallization process in the liquid phase supercooled  below T$_m$ and in amorphous phase overheated above T$_g$ by evaluating the crystal growth velocity ($v_g$) as a function of temperature in the range 500-940 K.  

The NN potential allowed us to perform simulations with several thousands of atoms for overall 100 ns  that provided  the crystal growth velocity $v_g$ as a function of temperature with greater details than reported previously by DFT simulations.
To this end, we first extended to a wider range of temperatures the simulations of the 12960-atom model of the liquid-crystal interface discussed in previous sections. 
The model was first quenched in 40-80 ps  from 900 K to each  target temperature to monitor the evolution of the crystalline slab.
The number of crystalline atoms is quantified by using the local order parameter for crystallinity  Q$_4^{dot}$ 
\cite{q4} that is suitable  to distinguish atoms in the crystalline phase from atoms in  liquid/amorphous environments as shown in the Supplementary Figure 10.
Then, the evolution of the crystalline interface was monitored to estimate the crystal growth velocity as  $v_g=dL(t)/dt$  where $L(t)$ is the effective (half) thickness of the crystalline slab given by $L\left(t\right)=N(t)/{2A\rho_{NN}^{cubic}}$, where $N(t)$ is the number of atoms in the crystalline slab, $\rho_{NN}^{cubic}$ is the theoretical equilibrium density of the cubic phase, $A$=6.15 $\times$ 6.15 nm$^2$ is the cross-section of the cell orthogonal to the growth direction,  and the factor two at the denominator accounts for the presence of two growing surfaces. The evolution of $L(t)$ as a function of time at several temperatures is reported in  the Supplementary Figure 11. Since the crystal-liquid interface lays on the (001)
plane of the cubic phase, the crystal growth velocity corresponds to the growth along the [001] direction of the cubic crystal.
 Before analyzing the results, we verified that our thermostat was effective in getting rid of the latent heat of crystallization released during crystal growth that in the real system diffuses away very fast due to electronic thermal conductivity of the liquid. To this aim, we  considered slices (bins) 10 \AA\ wide at different distances from the liquid-crystal interface as shown in the Supplementary Figure 12. The local temperature  and the fraction of crystalline atoms in the different slices are shown in the Supplementary Figures  13-17 for different average temperatures. The local temperature is indeed rather uniform across the liquid-like slab at different distances from the surface.
The crystal growth velocity as a function of temperature is often  described by the phenomenological  Wilson–Frenkel formula (WF) \cite{WF1,FrenkelJ} $v_g = u_{kin}(1 - exp(-\Delta\mu/k_BT))$  where $u_{kin}$ is a kinetic prefactor and $\Delta\mu$ is the free energy difference between the crystalline and supercooled liquid phases. 

 For a diffusion-controlled growth, the kinetic prefactor is given by $u_{kin}=6Ddf/\lambda^2$, where $\lambda$ is the typical jump distance of atoms in the elementary diffusion process, $D$ is the diffusion coefficient, $d$ is the interlayer spacing along the growth direction and $f$ represents the fraction of surface sites  where a new atom can be incorporated \cite{binder}.
The WF formula is typically adequate to describe a continuous growth of  a rough surface as it seems to be the case here; a snapshot of the growing surface is given  in the Supplementary Figure 18.

By setting $d=\lambda$ and $f$=1 the kinetic prefactor has the form $u_{kin}=\frac{6D}{\lambda}$ that we used for instance in a previous simulation of the crystallization of GeTe
from a crystal-liquid interface (heterogeneous crystallization) \cite{Sosso2015}. 
We spend a few words on the justification of the WF formula in view of possible different choices for the kinetic prefactor that we will discuss later on.
The crystal growth velocity can be expressed as
$v_g = d f (\kappa^+-\kappa^-$), where $\kappa^+/\kappa^-$ are the rate of attachment/detachment of an atom to/from the crystalline surface. In turn $\kappa^+= \nu exp(-\Delta  G^*/k_BT)$ and $\kappa^-= \nu exp(-\Delta G^*/k_BT -\Delta   \mu / k_BT)$, where $\Delta G^*$  is the activation energy for the attachment to the surface  of an atom from the liquid with an attempt frequency $\nu$. By assuming that $\kappa^+$ is equal to the rate of a jump in 
the diffusion process of a single atom in the liquid, it can be written as  $\kappa^+=6D/\lambda^2$. This approximation leads to $v_g =6Ddf/\lambda^2 (1- exp(-\Delta \mu /k_B T))$. We remark, however, that $f$ should also include possible other 
corrections to the sticking.
Due to the uncertainties in the form of the kinetic prefactor,
$\lambda$ is typically considered as a fitting parameter.

\begin{figure*}[t]
 \centering
  \includegraphics[width=0.99\textwidth, keepaspectratio]{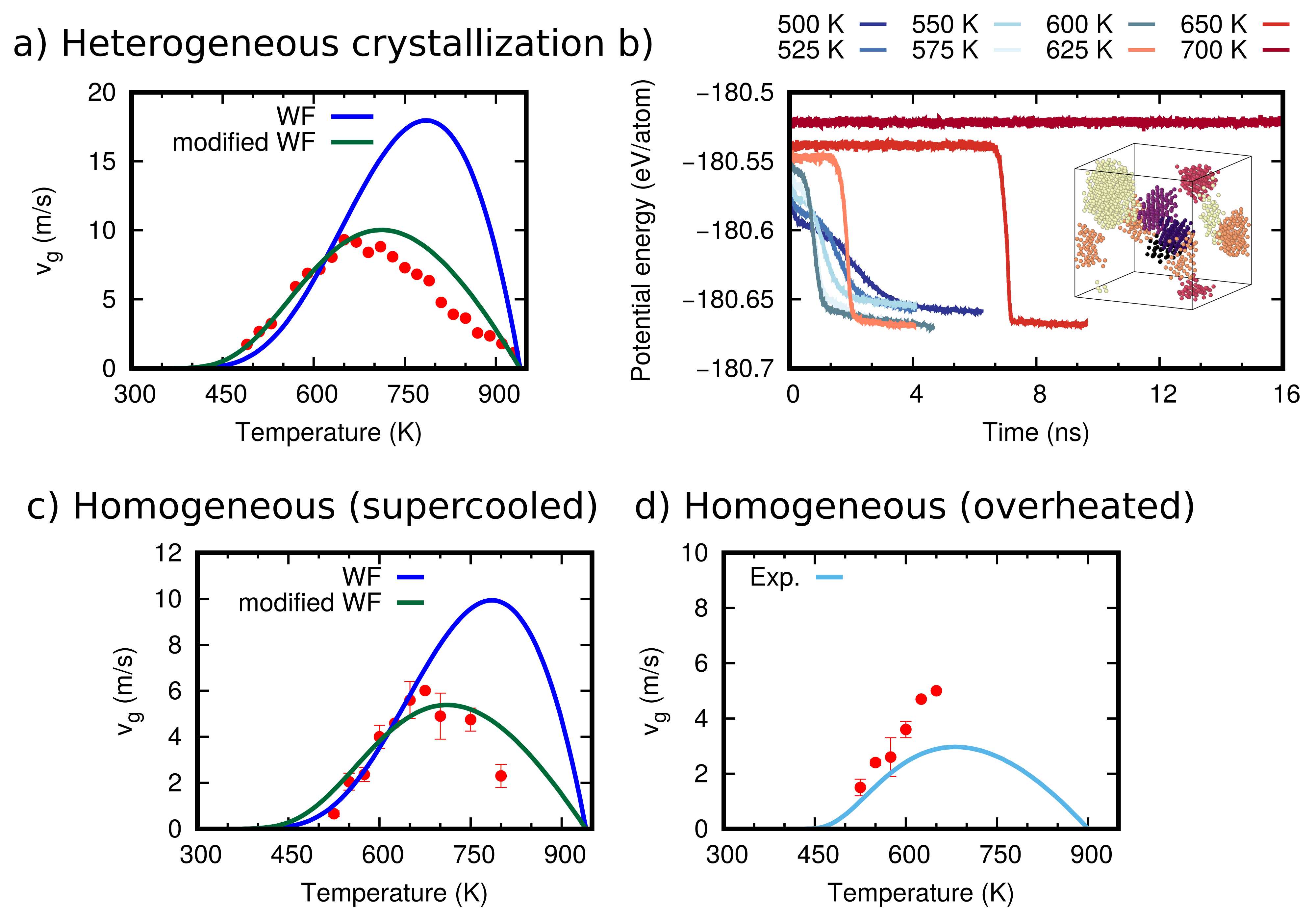}

 \caption{\textbf{Crystal growth velocity.} a) Crystal growth velocities ($v_g$) extracted from the motion of the crystal-liquid interface in the MD simulations (red dots).  The crystal growth is along the [001] direction of the cubic phase. The dark blue and green lines correspond to WF and modified-WF fits (see text).  b) Potential energy as a function of time in simulations of the homogeneous crystallization of the supercooled liquid phase at different temperatures. The inset shows a snapshot of the formation of several crystal nuclei (with different colors) at 500 K, liquid-like atoms are not shown. c)  Crystal growth velocities ($v_g$) for the homogeneous (red dots) crystallization. 
 The dark blue and green lines correspond to WF and modified-WF fits (see text).
  d) Crystal growth velocity (red dots) of a  model of the amorphous phase overheated at different temperatures above T$_g$ (see text). The light blue line refers to the  results of Ref. \cite{orava} inferred from DSC data available below 650 K.}
 \label{A3}
\end{figure*}

We attempted to reproduce the crystal growth velocity extracted from MD simulations with the WF relation and  $u_{kin}=\frac{6Dd}{\lambda^2}$ (i.e. $f$=1) with d=3.0 {\AA } for the growth along the [001] direction of the cubic phase, by using the theoretical $D$ obtained from the  CG fit of the MD data and by using for  $\Delta\mu$ the expression given by Thompson and Spaepen  $\Delta\mu(T)=\frac{\Delta H_{m}(T_m-T)}{T_m}\frac{2T}{T_m+T)}$ \cite{thompson1979approximation}. We set $\Delta H_{m}$ = 166 meV/atom, $T_m$ = 940 K as estimated from our MD simulation (see previous sections).  We also checked that the Thompson-Spaepen formula is fairly accurate in our case by computing $\Delta\mu$ from the integration of the specific heat as $\Delta\mu(T)=\Delta H_{m}(1-\frac{T}{T_m})-\int^{T_m}_T\Delta C_pdT+T\int^{T_m}_T\frac{\Delta C_p}{T}dT$,
where $\Delta C_p$ is the difference in $C_p$ between the amorphous and the crystalline phases. We approximated $\Delta C_p$ with $\Delta C_v$ which was in turn computed from the caloric curve at fixed density as discussed in Sec. 2. The resulting $\Delta\mu$ in the temperature range of interest is nearly indistinguishable from that obtained from the Thompson-Spaepan approximation as shown  in the Supplementary Figure 19. 
 Actually, we have not been able to reasonably fit the data over all temperatures 
by using just $\lambda$ as a free parameter. Therefore we restricted the fit to the lower temperatures (below 650 K) which yields  the  WF curve shown in Fig. \ref{A3}a with a physically reasonable value of $\lambda$= 2.42 {\AA }.  
The WF curve reproduces the  crystal growth velocity at low temperatures but largely overestimates $v_g$ at high temperatures. 
  We first discuss the behavior at low temperatures below 650 K. The good fitting with the WF formula means that $u_{kin}$ is thermally activated with an activation energy close to that of the self-diffusion coefficient. 
We remark that the self-diffusion coefficient describes the long-scale atomic diffusion process of individual atom while the term that enters in the kinetic prefactor $u_{kin}$ for  crystallization is actually an effective  diffusion coefficient $D_{eff}$ that might embody a more short-ranged atomic motion. 
In glasses close to T$_g$, the secondary $\beta$-relaxation is known to be the dominant source of atomic dynamics while the slower $\alpha$ relaxation controls the long range atomic diffusivity \cite{cavagna}. At high temperature far from T$_g$, the $\alpha$ and $\beta$ relaxations actually coincide \cite{cavagna}. 
The possibility that $\beta$-relaxation might enhance the crystallization kinetics of phase change materials close to T$_g$ has been put forward very recently \cite{DSM,WuttigGeSb}.
Indeed, the presence of a  $\beta$-relaxation process
in phase change materials has been recently identified from
the so called $\beta$-wing in the temperature dependence of the loss modulus measured by dynamical mechanical spectroscopy (DSM) \cite{DSM}.
Evidences of a link between the crystallization speed and the presence of $\beta$ relaxation have been provided very recently for the eutectic alloy Ge$_{15}$Sb$_{85}$ \cite{WuttigGeSb}.
Molecular dynamics simulations have also revealed that the stabilization of the amorphous phase of Sb in ultrathin films is possibly due to the reduction of the $\beta$-relaxation dynamics due to the confinement \cite{dragoni2021mechanism}. 
It is therefore of interest to investigate whether $\beta$-relaxation might also  be of relevance for crystal growth in GST225. The $\beta$-relaxation  can be detected in MD simulations by looking at the intermediate scattering function \cite{cavagna}, as we did for instance for  the GeTe phase change compound in our previous work \cite{sossoDH}.  There, we also used four-point correlation functions \cite{donati} and isoconfigurational analysis \cite{Widmer-Cooper} to investigate dynamical heterogeneities in the supercooled liquid phase which is associated with the fragility and the breakdown of the Stokes-Einstein relation between $D$ and the viscosity. The availability of a NN potential for GST225 now allows for an extension of our previous analysis on GeTe to the ternary compound. However, we leave this new  chapter on the properties of the supercooled liquid phase for a future work as the present one is already very dense of information.
Therefore, here the $\beta$-relaxation is addressed in a simpler manner by looking  at the MSD as a function of time and at different temperatures.
 A plateau in the MSD plotted in a log-log scale close to T$_g$ is typical of a two steps relaxation dynamics with a faster $\beta$-relaxation and a slower $\alpha$-relaxation that controls the long range atomic diffusivity after the plateau \cite{cavagna}.  In GST225, a clear plateau is present at 400 K, although an inflection starts to appear in the MSD at 500 K which is the lowest temperature at which we have investigated the crystallization kinetics (see Supplementary Figure 20). This suggests that the $\beta$-relaxation dynamics starts to appear only very close to T$_g$  which is consistent with the presence of a $\beta$-wing in the experimental DSM data only below 440 K \cite{DSM}.
{ We mention that previous DFT molecular dynamics simulations 
suggested that the fast crystallization in GST225 is characterized by concerted atomic motions favored by  the presence of flexible axial bonds \cite{elliott2017}.  Our results on the crystal growth velocity suggest that these atomic motions would still feature an activated behavior with an activation energy close to that of the self-diffusion coefficient.

 Turning now to the crystal growth velocity above 650 K, it is clear from Fig.  \ref{A3}a that the WF formula is unable to reproduce the data from the simulations. The behaviour at high temperature is particularly sensitive to the value of $\Delta\mu$ which goes to zero at T$_m$.
This misfit could therefore be somehow reduced by changing $\Delta H_{m}$ in the range 0.12-0.24 eV/atom and T$_m$ in the range 860-940 K due to the uncertainties in these figures discussed in previous sections. The results for the crystal growth velocities obtained from the WF formula and different values
of  $\Delta H_{m}$  and T$_m$ are reported in the Supplementary Figure 21. Still a sizable disagreement between the WF formula and the crystal growth velocities extracted from the simulations is present even for the best choice of $\Delta H_{m}$ and T$_m$ in the ranges given above.

 There are actually different examples in metals and semiconductors with diffusion-controlled crystallization kinetics in which the WF formula does not quantitatively predict results from simulations or experiments \cite{binder,reed,stolk}. This discrepancy has been ascribed to changes in the mobility of the supercooled liquid in the proximity of the crystal interface \cite{binder}.
To assess the origin of this discrepancy in GST225, we have thus  analyzed the local atomic mobility as a function of the distance from the crystal-liquid interface. The local 2D diffusion coefficient in obtained from the Einstein relation 
$<x^2> + <y^2>$=4$D$t in NVT simulations at different average temperatures where the WF formula fails. The $x$
and $y$ directions lye in the interface plane. The local $D$ is computed  in the slices at different distances from the interface shown in  the Supplementary  Figure 12. 
The local $D$ reported in  the Supplementary  Figure 22 is indeed lower at the interface than deep in the liquid as also discussed in Ref. \cite{binder}.
However, the reduction in the mobility closer to the interface  is not sufficiently large to justify the overestimation of  $v_g$ given by the WF formula once  the bulk value of $D$ is used.

In the attempt to fit the $v_g$ data at high temperatures, we reconsidered the form of $u_{kin}$ following the prescription by Jackson \cite{jackson} who multiplied $\kappa^+$ and $\kappa^-$  by a factor $e^{-\Delta S/k_B T}$ where $\Delta S$ is the (positive) entropy difference between the liquid  and the crystal.
This expression, which is very seldom used \cite{pichon,asta,broughton}, was justified by considering that $e^{-\Delta S/k_B T}$ is the ratio of the number configurations in the crystal and in the liquid and that the rate at which an atom can be incorporated in the crystal depends on the rate
at which an ergodic sampling of configurations in the liquid would find a crystalline configuration
\cite{harrowell}.

At low T, $\Delta S$ is small and then the correction has a minor effect on $v_g$, while $\Delta S \approx 2.1 k_B$ at T$_m$  which leads to a large reduction of the crystal growth velocity (see Supplementary Figure 23).
$\Delta S$ as a function of temperature was obtained as $\Delta S=\Delta H_m/T_m - \int^{T_m}_T\frac{\Delta C_v}{T}dT$, with $\Delta C_v$ given by the caloric curves in Fig. \ref{fig:diff}c.
The modified WF formula $v_g =6Dd/\lambda^2 e^{-\Delta S/k_BT} (1- exp(-\Delta \mu/k_BT))$ turns out to fit reasonably well the data at all temperatures as shown in Fig. \ref{A3}a with $\lambda$=1.65 {\AA } which is still a reasonable number for the typical jumping distance. The overall small misfit at high temperatures could now be accounted for by the slightly lower mobility close to the surface that we discussed above (see  Supplementary Figure 22).
As a final remark, we also mention that Kelton and Greer \cite{kelton_greer} proposed the different expression for the attachment/detachment rates $\kappa^+= \nu exp(-\Delta   G^*/k_BT +\Delta \mu / 2k_BT)$, $\kappa^-= \nu exp(-\Delta G^*/k_BT -\Delta \mu / 2k_BT)$ which leads to $v_g =6Ddf/\lambda^2 (1- exp(-\Delta \mu/k_B T)) e^{\Delta \mu/2k_B T}$. The fit of the data with this latter formula is, however, less satisfactorily, as shown in the Supplementary Figure 24.

We also mention that the formula for $v_g$ given by the two-dimensional surface nucleation growth model \cite{jackson2} does not work neither for reasonable values of the crystal/liquid interface energy in the range 0.05-1 J/m$^2$ \cite{kohari} which enters as a parameter in the formula (see for instance Eq. 5 in Ref. \cite{nascimento}), which is consistent with the mostly continuous growth behavior observed for GST225.}

 We have also repeated the simulation of the crystallization from the crystal-liquid interface along the [111]
direction of the cubic phase. The results at different temperatures are compared with those of the (100) surface in   the Supplementary Figure 25. The crystal growth velocity at the (100) and (111) surfaces are essentially the same, which is somehow surprising given that in simulations of GeTe  the crystal growth velocity was  significantly lower for the  (111) than for the (100) surface \cite{Sosso2015}. This behavior in GST225 might be ascribed to the high roughness of the growing surfaces
(see Supplementary Figure 18).

We have also studied the crystallization in the bulk (homogeneous crystallization) of the supercooled liquid 
 to directly compare our results with the experimental DSC data in Ref. \cite{orava} that refer to these conditions. To this aim, we generated a 7992-atom cubic model of the liquid phase at 990 K as discussed previously. Then, we quenched the model down to different temperatures in 80 ps MD simulations in the NVT ensemble and at the experimental density of the amorphous phase (0.0309 atom/\AA$^3$). Finally, we performed long simulations up to 20 ns to crystallize the supercooled liquid and to extract the crystal growth velocity.  The  potential energy as a function of time for simulations at different temperatures shown in  Fig. \ref{A3}b reveals the onset of the crystallization with an incubation time that increases with temperature. Overcritical nucleus/nuclei form on a time scale of 0.2 to 3 ns in the range 500-600 K, and after about 7 ns  at 650 K. Nucleation was not observed at and above 700 K in simulations lasting over 20 ns. For temperatures where nucleation was not observed after a few tens of nanoseconds, the crystal growth velocities were estimated by heating at the target temperature a configuration  with an overcritical nucleus generated at a lower temperature. Below 600 K the number of overcritical nuclei increases as temperature decreases (see inset of Fig. \ref{A3}b). For temperatures above 650 K, we observe a single crystallite which gives rise to a uniform single crystal filling the simulation box. Crystalline atoms are assigned to a crystalline nucleus when they fall within a 3.6 \AA\ from the outermost atoms of the nucleus.  The critical nucleus increases in size with temperature and it contains about 40 atoms at 500 K and 50 atoms at 650 K.

The evolution of the crystalline nuclei was monitored to evaluate $v_g$ which for a spherical nucleus with radius $R$ is given by  $v_g={dR(t)/dt}$ with $R\left(t\right)=\left({3N(t)}/{4\pi\rho_{NN}^{cubic}}\right)^{\frac{1}{3}}$, where $N$ is the number of atoms in the  nucleus.  This assumption is valid only in the early stage of crystallization when  the nuclei do not interact with each other  or with their periodic image. 
The evolution of $R(t)$ of the nuclei  at several temperatures is reported  in the Supplementary  Figure 26.  The resulting $v_g$, averaged over different nuclei, are shown in Fig. \ref{A3}c. 

\begin{figure*}[ht]
 \centering
 \includegraphics[width=\textwidth, keepaspectratio]{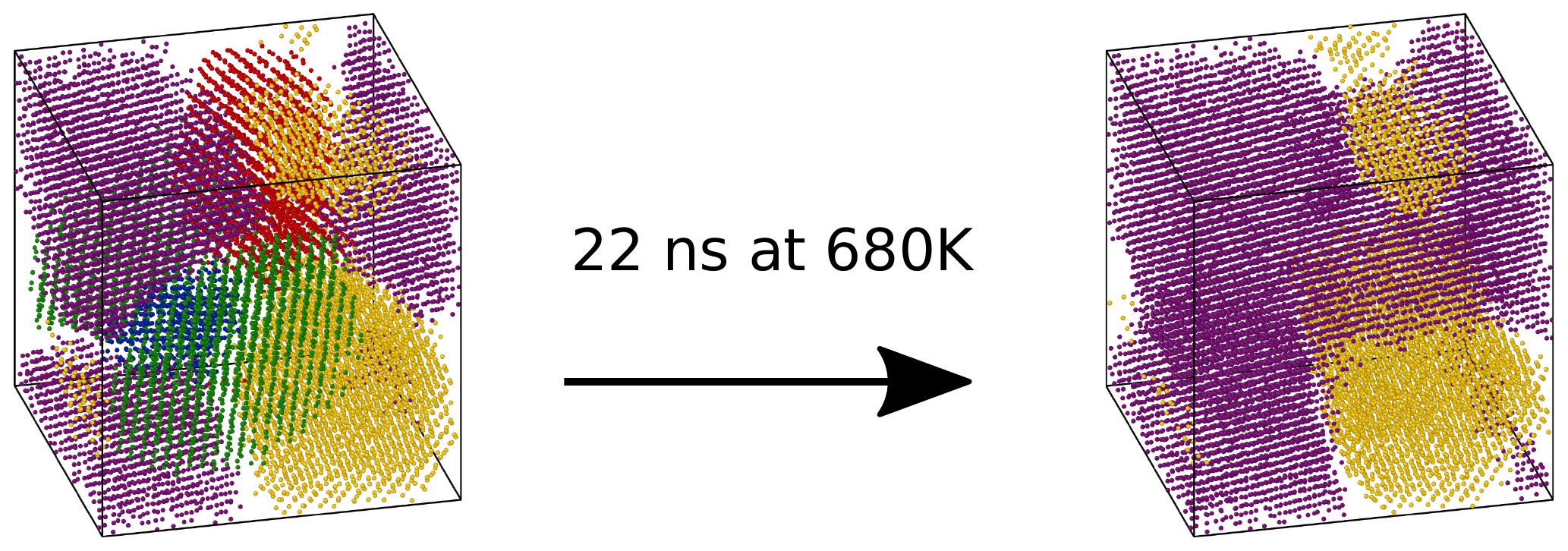}

 \caption{\textbf{Coarsening.}  Simulation of crystallites coarsening in a 27000-atom supercell at 680 K. 
 Several nuclei, shown by different colors in the left panel, coalesce after 22 ns into only two crystallites shown in the right panel.}
 \label{coarsening}
\end{figure*}

 Notice that the heterogeneous crystal growth velocity is higher than the homogeneous one as it was the case for GeTe \cite{sosso2013,Sosso2015}. This difference can partially be accounted for by a geometric factor. The interlayer distance $d$ entering in the WF formula for the crystal growth velocity at the crystal/liquid interface  turns for a spherical nucleus into the
factor $4/3 (vol_{site}3 / (4 \pi))^{1/3} $, where $vol_{site}$ is the volume associated with an adsorption site on the crystalline nucleus \cite{raoux-ielmini}. If we take $vol_{site}=4 \pi /3 (\lambda/2)^3$, the WF formula reads $v_g =4D/\lambda (1- exp(-\Delta \mu/k_B T))$ that we used for GeTe with $\lambda$= 3.0 \AA \cite{sosso2013}. 
For GST225, we attempted to fit the data below 650 K with the more general formula  $v_g= 8 (vol_{site}3 / (4 \pi))^{1/3} D/\lambda^2 (1 - exp(-\Delta \mu /k_BT))$, where $vol_{site}^{1/3}$
is about half the lattice parameter of the cubic cell, i.e 3 \AA.
As expected, the resulting curve with $\lambda$= 3.02 \AA, shown in Fig. \ref{A3}c, overestimates the crystal growth velocity at high temperature.
A better fit is obtained by adding the entropic factor $e^{-\Delta S/k_BT}$, similarly to the heterogeneous case, as shown in Fig.\ref{A3}c for $\lambda$= 2.05 \AA. The difference in the values of $\lambda$ obtained from the best fit for the homogeneous and heterogeneous case is due to the fact that the geometric factor discussed above is in both case an approximated form because the surface is rough and the nucleus is not spherical. Moreover, the sticking  is not expected to be the same for an extended surface and for a small nanoparticle.

Finally, we repeated the simulations of the homogeneous crystallization for  the amorphous phase   overheated above T$_g$ in a 7992-atom cubic cell.  The amorphous model, first equilibrated at 300 K, was heated at different target temperatures in the range 500-800 K with a heating rate of 5 K/ps.  Note that the fraction of Ge atoms in tetrahedral configurations decreases very rapidly with temperature above T$_g$ and before crystal nucleation sets in, as shown by the distribution of $q$ order parameter reported in the Supplementary  Figure 27. Crystal nucleation was observed only below 650 K in the simulation time of 6 ns. The evolution in time of the radius of the nuclei  at different temperatures is reported in  in the Supplementary Figure 28.  
The resulting $v_g$ as a function of temperature for the overheated amorphous phase is compared in Fig. \ref{A3}d  with experimental DSC data from Ref. \cite{orava}. 
The crystal growth velocities for the overheated amorphous phase are very similar to those obtained for the supercooled liquid at the same temperature and they are very close to the experimental data from ultrafast DSC \cite{orava}.
At 600 K, for instance, 
 the crystal growth velocity from NN-MD is 4.0  m/s in the homogeneous  crystallization of the supercooled liquid, to be compared with the experimental value of about 2.5 m/s \cite{orava}. Other values obtained from DFT-MD simulations in literature at the same temperature  range from 5.5 m/s in small models, to 0.3-1 m/s for intermediate models and 0.5 m/s for the largest models \cite{elliot,PhysRevLett.107.145702, kalikkacrys3,ronneberger2015crystallization,ronneberger2018crystal,PhysRevB.84.094124,kalikkacrys1,kalikkacrys2}.\\
  Just to provide a last example of the capability of the NN potential, we show in Fig. \ref{coarsening}
 the evolution in time of a 27.000-atom model at 680 K in which we observed the coarsening of crystalline nuclei on the time scale of 20 ns.

\section{DISCUSSION}

In summary, we have devised a machine learning neural network interatomic potential for GST225 based on the DeePMD scheme 
 which employs deep neural networks to fit a large DFT database
\cite{wang2018deepmd}. The NN potential is highly accurate as it reproduces the DFT results for a wide range of structural, dynamical and thermodynamical properties of the crystalline, amorphous and liquid phases. The NN potential has been exploited in  large scale (12000 atoms for over 100 ns) MD simulations of the supercooled liquid and overheated amorphous phases which yielded the crystal growth velocities in a wide  temperature range of interest for the operation of the memory devices in good agreement with experimental data from ultrafast DSC \cite{orava}.  
 In the temperature range 500-650 K, the crystal growth velocity extracted from the simulations show an activated behavior controlled by the self-diffusion coefficient. The  analysis of the MSD showed that the $\beta$-relaxation seems to be present only at temperatures very close to T$_g$.  A modified form \cite{jackson} of the WF formula turns out to be suitable to describe the crystal growth velocity on a wider temperature range from T$_g$ to T$_m$.

We remark that the computational cost of the NN potential scales linearly with the system size  \cite{lu202186} which would allow simulating the entire volume of the active material of ultrascaled memories (linear scale of 10 nm)  on the timescale of the device operation. This would allow addressing several issues such as crystallites coarsening and evolution of grain boundaries which are particularly relevant for multiscale programming exploited in the neuromorphing computing, just to name a few others besides those already mentioned in the introduction.  Extension of the potential to Ge-rich GeSbTe alloys would also allow uncovering the mechanism of phase separation into Ge and less Ge-rich ternary alloys which is believed to be responsible for the raise of the crystallization temperature upon Ge enrichment  of interest for  applications  in  memories embedded in microcontrollers \cite{Cappelletti_2020}.
\vskip 1.0 truecm
\section{METHODS}
\vskip 0.5 truecm
 The NN potential was obtained by fitting DFT  energies, forces and the stress tensor of a database containing about 180000 supercell models (configurations) of GST225 in the liquid, amorphous, cubic and hexagonal phases by  using the DeePMD-kit open-source package \cite{wang2018deepmd,PhysRevLett.120.143001}. 

DFT calculations were performed by using the Perdew-Burke-Ernzherof (PBE) exchange and correlation functional \cite{PBE} and Goedecker-Teter-Hutter (GTH) norm conserving  pseudopotentials with $s$ and $p$ valence electrons \cite{GTH1,GTH2} as implemented in the CP2k package \cite{CP2k}. Kohn-Sham orbitals were expanded in Gaussian-type orbitals of a triple-zeta-valence plus polarization basis set, while a basis set of plane waves up to a kinetic energy cutoff of 100 Ry is used to represent the charge density as implemented in the Quickstep scheme \cite{CP2k}. This DFT-PBE framework has been shown in previous works to reproduce well the experimental structural and dynamical   properties of GST225 in the liquid, amorphous and crystalline phases \cite{caravati2007coexistence,caravati2009first,BPMGST,campi,baratella2022,cobelli2021}.\\
The liquid and amorphous phases were modeled by a 108-atom cubic supercell. The cubic metastable phase was modeled by a cubic 57-atom  and by an ortorhombic 98-atom supercells. The hexagonal phase was modeled by a 108-atom supercell. 
To generate the database configurations,  DFT-MD simulations were performed in the Born-Oppheneimer approximation with a timestep of 2 fs and by restricting  the Brillouin Zone (BZ)  integration to the supercell $\Gamma$-point. Then, for a subset of the atomic configurations extracted from the MD trajectories, energy and forces were computed  by properly integrating the BZ with a higher cutoff of 400 Ry and then were used as a training set for the NN. Integration of the Brillouin Zone was performed over a 3x3x3  k-point mesh for 108-atom cell or a 4x4x4 mesh for smaller cells.\\
In the construction of the NN potential, we employed  two-body   and  three-body embedding descriptors (see  Ref. \cite{lu202186,DeepmdD3}) by including the third or the first coordination shell, respectively. To this end, we set the distance cutoff to 7 \AA\ for the two-body descriptor and to 3.8 \AA\ for the three-body descriptor.

We set the maximum number of neighbors to 30, 30 and 40 for Ge, Sb and Te for the two-body descriptor and to 10 for the three-body descriptors for all species. The embedding network has 3 hidden layers with 20, 40 and 80 neurons for the two-body descriptors, and 3 hidden layers with 3, 6 and 12 neurons for the three-body descriptors. Finally, the  network for the fitting of energy and forces  consists of 4 hidden layers with 120, 60, 30 and 15 neurons with atomic reference for the chemical species of -102.297, -146.521 and -218.984 eV for Ge, Sb and Te atoms, obtained from isolated atoms calculations with CP2k. In the embedding and fitting network, we have used the hyperbolic tangent as an activation function. We have also exploited the residual neurons as discussed in Ref. \cite{resnet}. The hyperparameters which control the learning process according to Ref. \cite{wang2018deepmd} are reported in the Supplementary Table 4.
The NN-MD simulations were performed with the LAMMPS code \cite{LAMMPS} exploiting GPU acceleration with a timestep of 2 fs and a Nos\'e-Hoover thermostat \cite{noseart,hoover}. Finally, we exploited the Ovito\cite{ovito} tool for the visualisation and the generation of all atomic snapshots of this manuscript. \\

\textbf{Data availability} \par The NN potential,  the training DFT database and atomic trajectories of the crystallization process are available at https://doi.org/10.24435/materialscloud:a8-45 
\medskip

\textbf{Acknowledgements} \par %delete if not applicable))
The project has received funding from European
Union NextGenerationEU through the Italian Ministry of
University and Research under PNRR M4C2I1.4 ICSC 
Centro Nazionale di Ricerca in High Performance Computing,
Big Data and Quantum Computing (Grant No.
CN00000013)

\medskip

\textbf{Code availability} \par
LAMMPS,  and DeePMD are free and open source codes available at https:// lammps.sandia.gov  and http://www.deepmd.org, respectively.
\medskip

\textbf{Competing interests} \par
The authors declare no competing interests.
\medskip

\textbf{Authors contributions} \par
O.A.E.K. and M.B. designed the research, O.A.E.K. and M.B. performed the research and analyzed data with the help of L.B. and M.P. for the generation of the NN potential. O.A.E.K. and M.B. wrote the paper and all the authors edited the manuscript before submission.
\medskip
\bibliography{ref1.bib}

\end{document}

\justifying

\captionsetup{justification=Justified}

\title{Unraveling the Crystallization Kinetics of the Ge$_2$Sb$_2$Te$_5$ Phase Change Compound with a  Machine-Learned Interatomic Potential  \\ SUPPORTING INFORMATION}

\author{Omar Abou El Kheir}
\affiliation{Department of Materials Science, University of Milano-Bicocca, Via R. Cozzi 55, I-20125 Milano, Italy}
\author{Luigi Bonati}
\affiliation{Atomistic Simulations, Italian Institute of Technology, Via Enrico Melen 83, 16152, Genova, Italy}
\author{Michele Parrinello}
\affiliation{Atomistic Simulations, Italian Institute of Technology, Via Enrico Melen 83, 16152, Genova, Italy}
\author{Marco Bernasconi}
\affiliation{Department of Materials Science, University of Milano-Bicocca, Via R. Cozzi 55, I-20125 Milano, Italy}

\maketitle

\newpage
\begin{table}[ht]
\centering
\captionof{table}{First maximum and minimum of the pair correlation functions in the liquid phase of GST225 at 990 K from NN-MD and DFT-MD (in parenthesis) simulations.}
\label{tab:gdrmax}

\begin{tabular}{c c c}
\hline
\hline
Bond type & First maximum (\AA) & First minimum (\AA) \\
\hline
GeGe& 2.71 (2.70)&3.28 (3.24)\\
GeSb& 2.90 (2.88)&3.44 (3.6)\\
GeTe& 2.82 (2.80)&-\\
SbSb&3.00 (3.00)&3.70 (3.78)\\
SbTe&3.05 (3.04)&-\\
TeTe&-&-\\
\hline
\hline
\end{tabular}
\end{table}

\begin{center}
\captionof{table}{Activation energy for the  self-diffusion coefficient in liquid GST225 from this work (NN-MD simulations) and previous works in literature.
NN+D2 refer to simulations with the semiempirical correction due to Grimme \onlinecite{D2} (D2) to include van der Waals interactions.}
\label{tab:DFF}

\begin{tabular}{c c c c c}
\hline
\hline
 E$_a$ (eV)& $D_0$ (10$^{-3}$cm$^2$ s$^{-1}$)& density (atom \AA$^{-3}$)& method & Reference \\
\hline
0.333&- &0.031 &PBE  &\onlinecite{caravatiDIFF}\\
0.300&- & -  &PBESol  &\onlinecite{MicoulautDIFF}\\
0.285&1.01&0.031 &NN+D2& This work\\
0.267&1.03&0.031 &NN&This work\\
0.260&0.45&0.031 &PBESol  &\onlinecite{elliotpssr}\\
\hline
\hline
\end{tabular}

\end{center}{}

\begin{figure}[h!]
 \centering
\includegraphics[width=\textwidth, keepaspectratio]{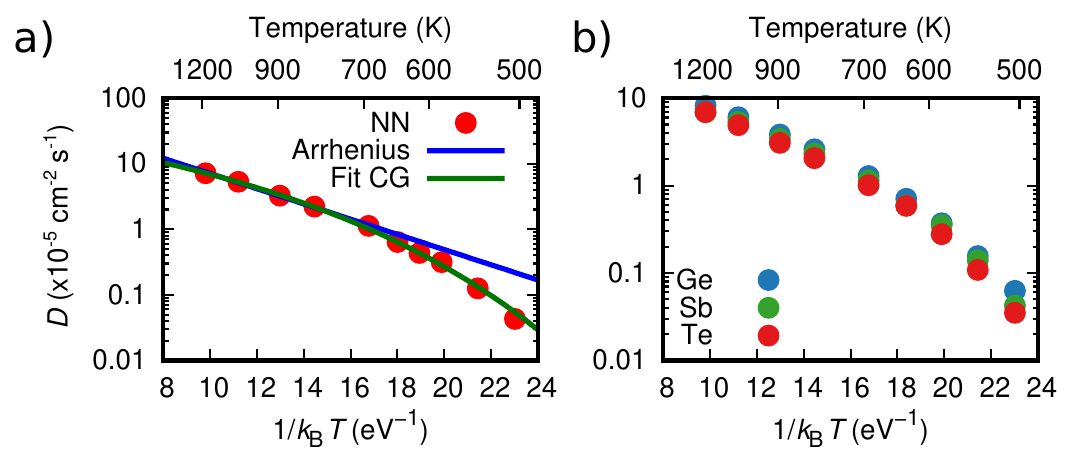}
 \caption{a) Cohen-Grest fit of the self-diffusion coefficient as a function of temperature (green curve) compared with the Arrhenius fit (blue curve, see Fig. 2d in the article).  b) The self-diffusion coefficient as a function of temperature resolved per chemical species.}
 \label{diff}
\end{figure}

\begin{figure}[h]
 \centering
 \includegraphics[width=\textwidth, keepaspectratio]{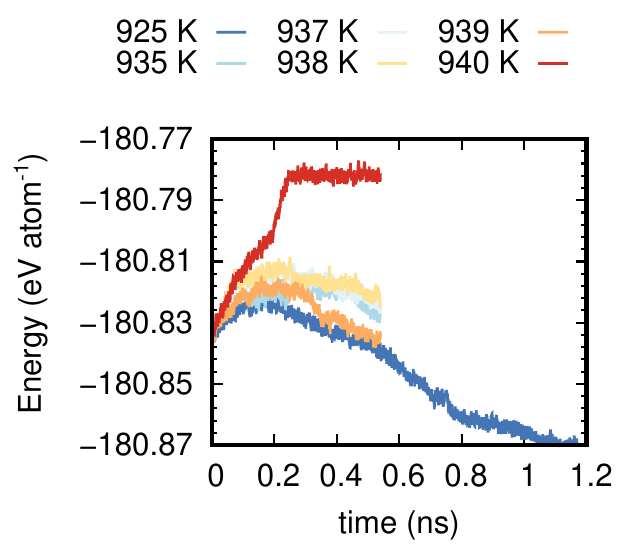}
 \caption{The potential energy as a function of time at different temperatures of the cubic-liquid interface model from NN+D2 simulations at constant (zero) pressure.  The melting temperature is estimated to be around 940 K.}
 \label{melting2}
\end{figure}
\newpage

\begin{figure}[h!]
 \centering
 \includegraphics[width=\textwidth, keepaspectratio]{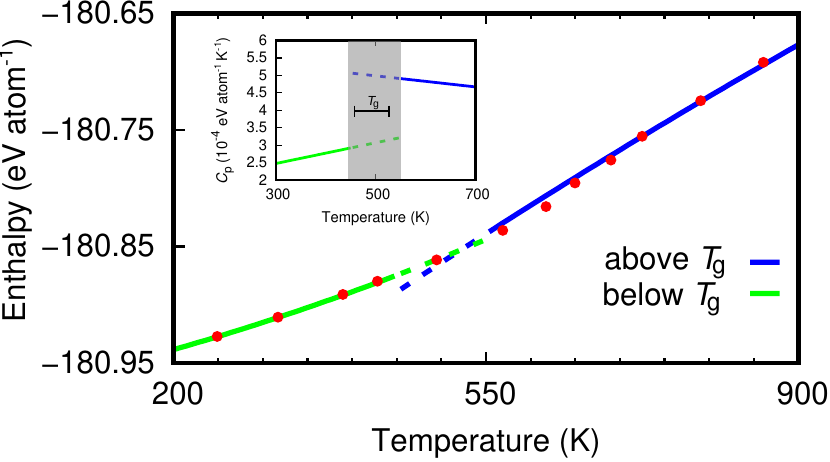}
 \caption{Enthalpy of the supercooled liquid as a function of temperature in NPT simulations. The lines are quadratic fit above and below $T_{\rm g}$. The heat capacity at constant pressure as a function of temperature obtained from the enthalpy fit is shown in the inset.}
 \end{figure}

\begin{table}[ht!]
\centering
\captionof{table}{Position (\AA) of the first maximum and minimum of the pair correlation functions in the amorphous phase of GST225 at 300 K from DFT and NN simulations. DFT data are given in parenthesis.  The Ge-Ge homopolar bond in the four DFT models generated for this benchmark is slightly longer (2.66$\pm$0.05 \AA) than that reported in Ref. \onlinecite{caravati2009first} (2.62 \AA) while it is shorter than that reported in Ref. \onlinecite{akola2008density} (2.84 \AA).}
\begin{tabular}{c c c}
\hline
\hline
Bond type & First maximum (\AA) & First minimum (\AA) \\
\hline
GeGe&2.62 (2.66)&3.14(3.15)\\
GeSb&2.82 (2.83)&3.6 (3.6)\\
GeTe&2.78 (2.78)&-\\
SbSb&2.99 (2.99)&-\\
SbTe&2.96 (2.97)&-\\
TeTe&3.04 (2.99)&-\\
\hline
\hline
\end{tabular}
\label{tab:agdrmax}

\end{table}

\begin{figure}[h!]
 \centering
  
 \includegraphics[width=\textwidth, keepaspectratio]{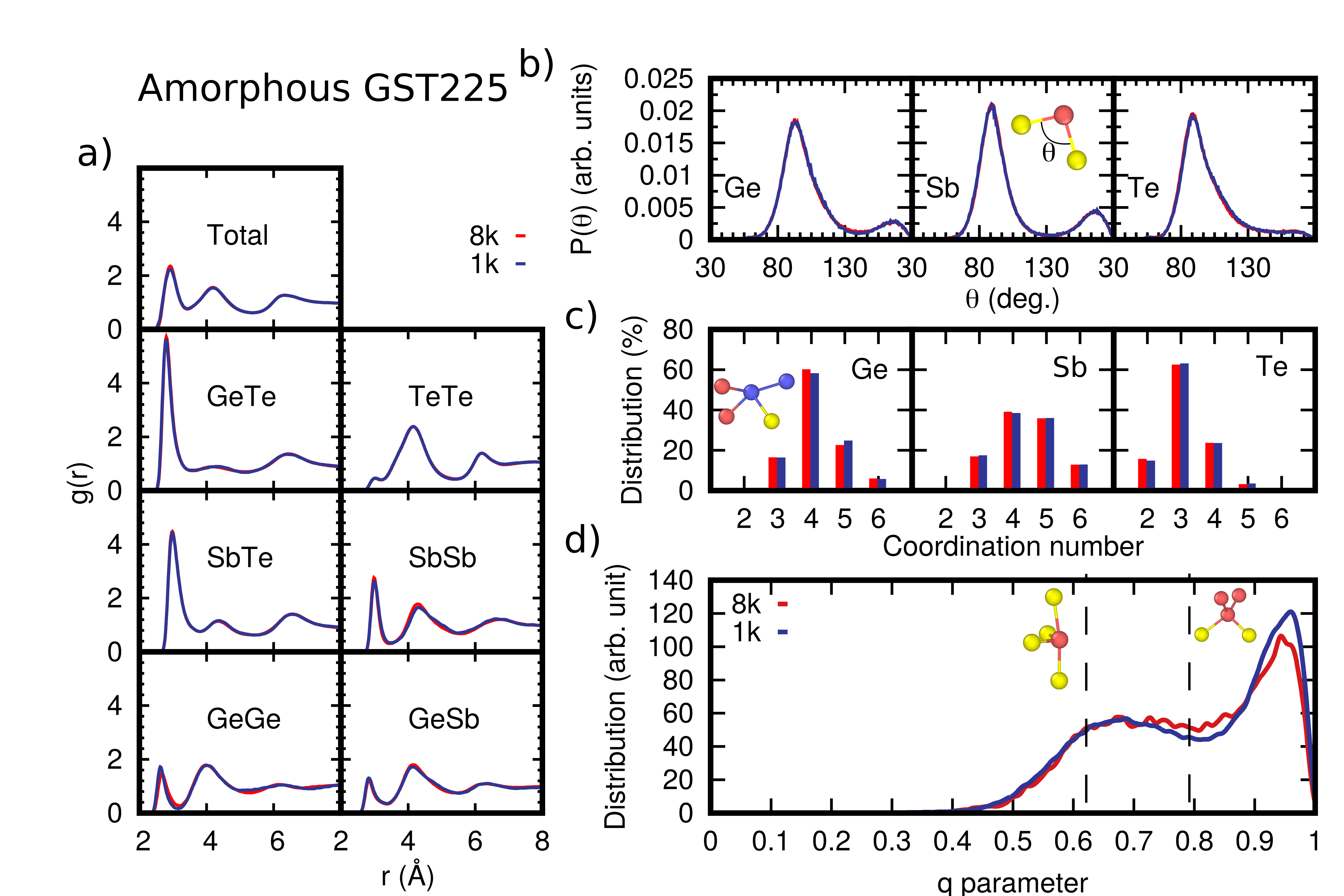}
 \caption{Structural properties of two NN models of amorphous GST225 at 300 K with 999 (blue curves) or 7992 (red curves) atoms.
 a) Total and partial pair correlation functions. b) Bond angle distribution functions resolved per central atomic species. The data are normalized to the number of triplets in each model. c) Distribution of coordination numbers resolved per chemical species. d) Distribution of the local order parameter $q$ for tetrahedricity for four-coordinated Ge atoms.}
\end{figure}

\newpage

\begin{figure}[h!]
 \centering
 \includegraphics[width=\textwidth, keepaspectratio]{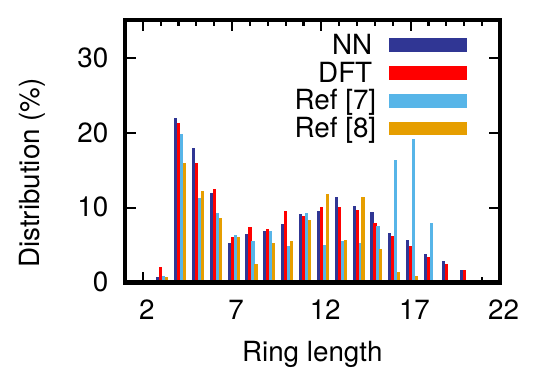}
 \caption{ Distribution of rings length  in amorphous GST225 at 300 K in the NN 999-atom  model and averaged over four DFT models, compared with previous DFT results  from Refs. \onlinecite{gabardi2012density} and \onlinecite{elliot}. The distribution is computed by using the algorithm in Ref. \onlinecite{RingAtomes}.}
\end{figure}

\clearpage

\begin{figure}[h!]
 \centering
 \includegraphics[width=\textwidth, keepaspectratio]{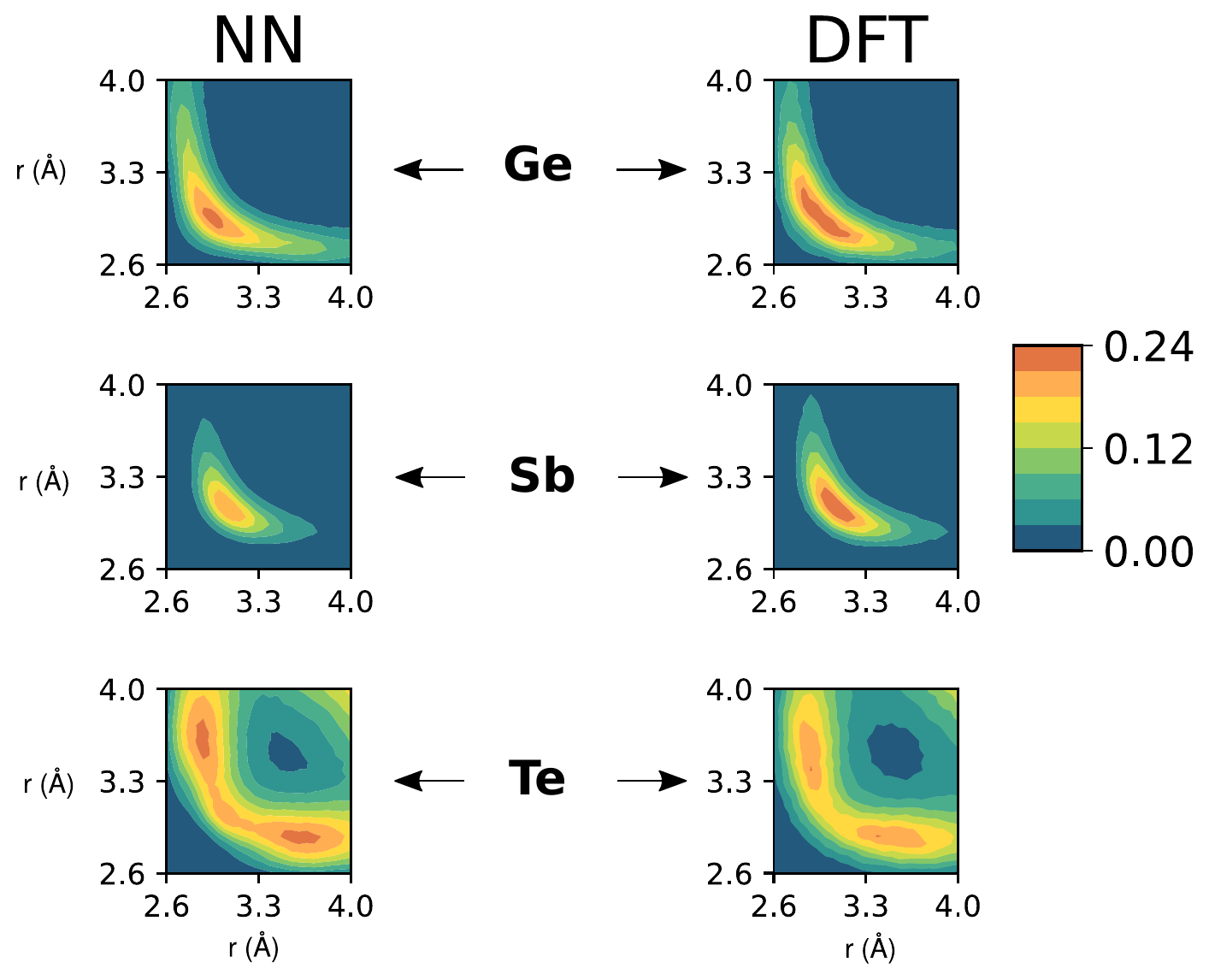}
 \caption{ALTBC distribution function of amorphous GST225 at 300 K resolved for each atomic species (Ge, Sb, and Te from top to bottom) in the NN model with 999 atoms (left) and  averaged over  four DFT models with 216 atoms (right). }

\end{figure}

\clearpage

\begin{figure}[h!]
 \centering
 \includegraphics[width=0.6\textwidth, keepaspectratio]{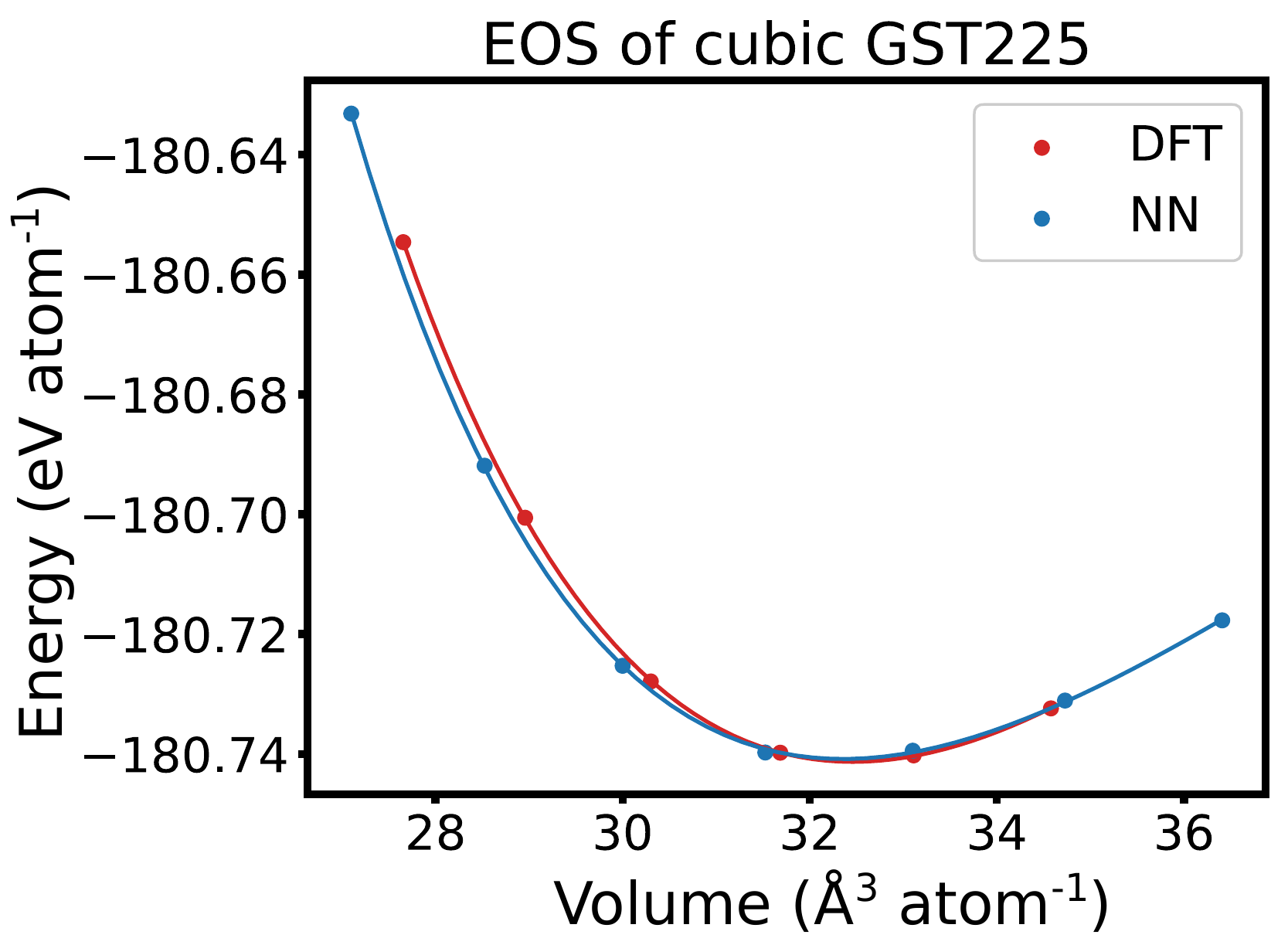}
 \caption{Equation of state of the cubic phase of GST225 from DFT and NN calculations.}
 \label{fig:EOScubic}
\end{figure}
\newpage
\begin{figure}[h!]
 \centering
 \includegraphics[width=0.6\textwidth, keepaspectratio]{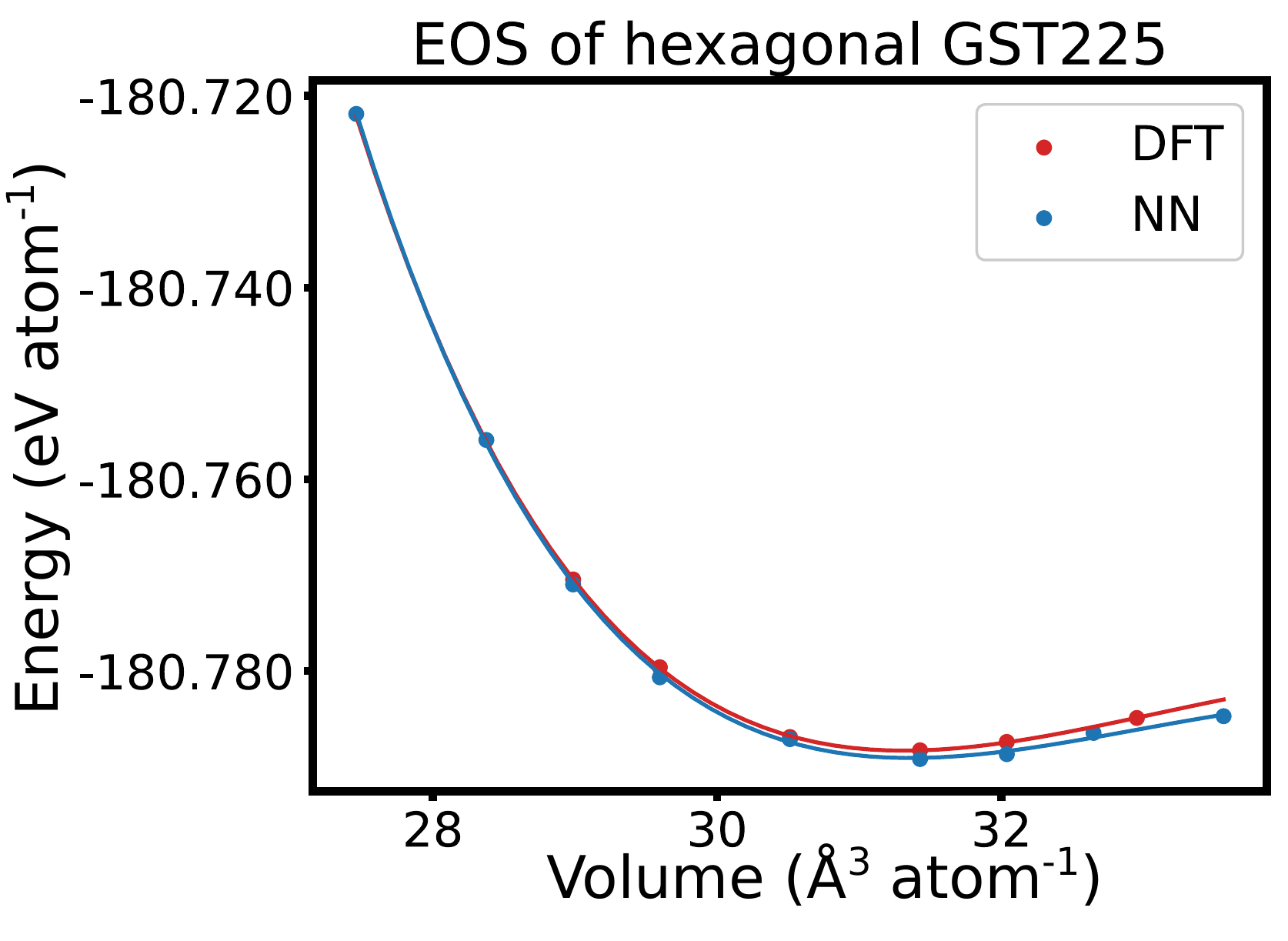}
 \caption{Equation of state of the hexagonal phase of GST225 (Kooi stacking, see article)  from DFT and NN
calculations.}
 \label{fig:EOSHex}
\end{figure}

\newpage

\begin{figure}[h!]
 \centering
 \includegraphics[width=0.8\textwidth, keepaspectratio]{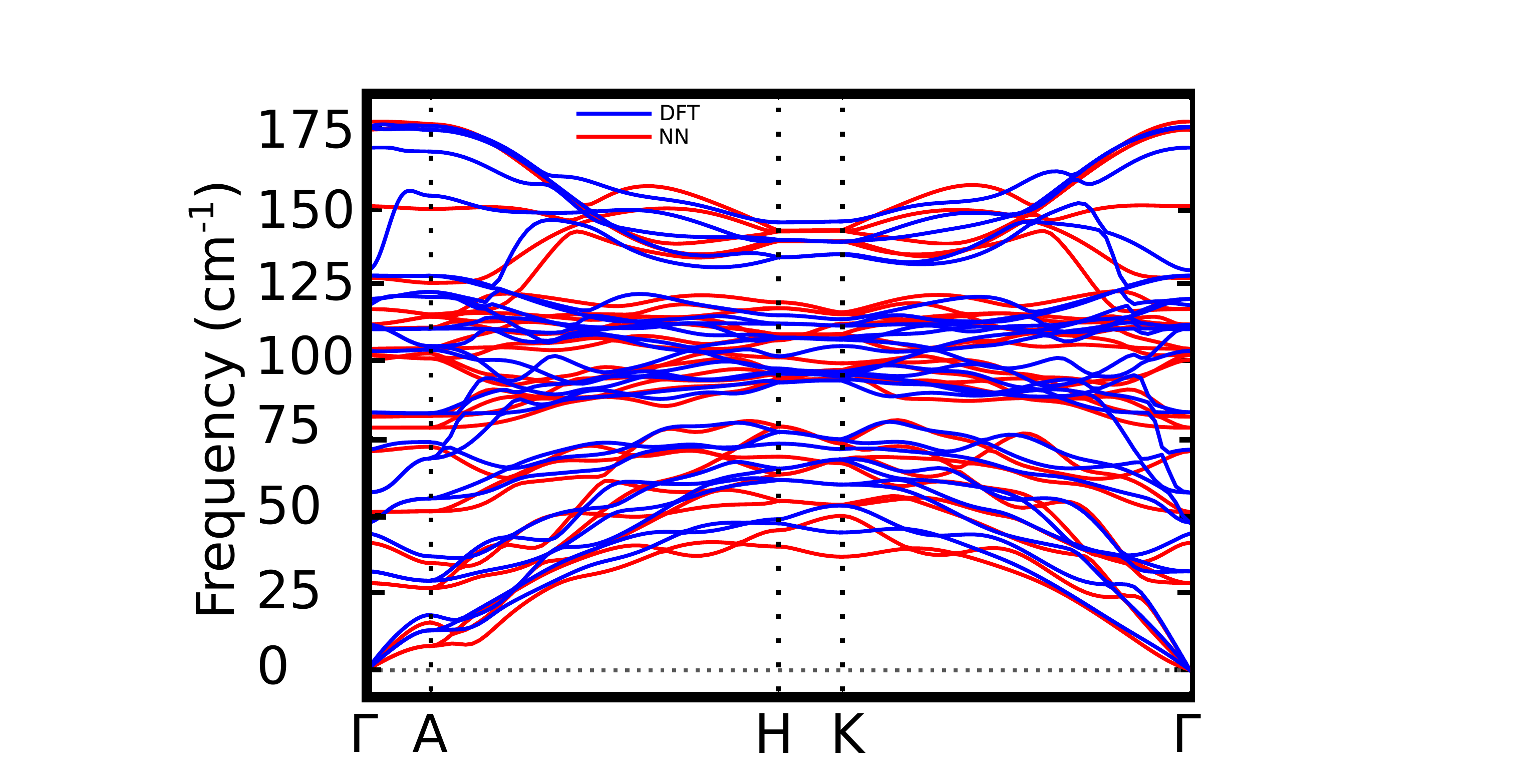}
 \caption{NN+D2 and DFT+D2 phonon dispersion relations of hexagonal GST225 (Kooi stacking, see article). DFT data are taken from Ref. \onlinecite{campi} and refer to a p-doped system with no long range Coulomb interactions. The derivatives of the dispersion curves close to the $\Gamma$-point provide the sound velocities that we averaged over all directions in the BZ to yield a transverse velocity of $v_{\rm t}$= 1270 m s$^{-1}$ and a longitudinal velocity of $v_{\rm l}$ = 2888 m s$^{-1}$. The DFT-D2 values are $v_{\rm t}$= 1950 m s$^{-1}$ and $v_{\rm l}$ = 3120 m s$^{-1}$ \onlinecite{campi}.
%From these values, we can also estimate the bulk thermal conductivity using the Cahill formula \onlinecite{cahill1992lower} for the minimum thermal conductivity given by $k_{min}=\left({\pi\rho^2}/{6}\right)^{\frac{1}{3}}\left(v_l+2v_t\right)k_B/2$
%where $\rho$ is the atomic density and k$_B$ is the Boltzmann constant. Within this approximation, our minimum thermal conductivity is estimated to be 0.3 W/m/K (0.43 from DFT+D2 data in Ref. \onlinecite{campi}). 
}
 \label{fig:branche}
\end{figure}

\begin{figure}[h]
  \centering
  \includegraphics[width=0.6\textwidth, keepaspectratio]{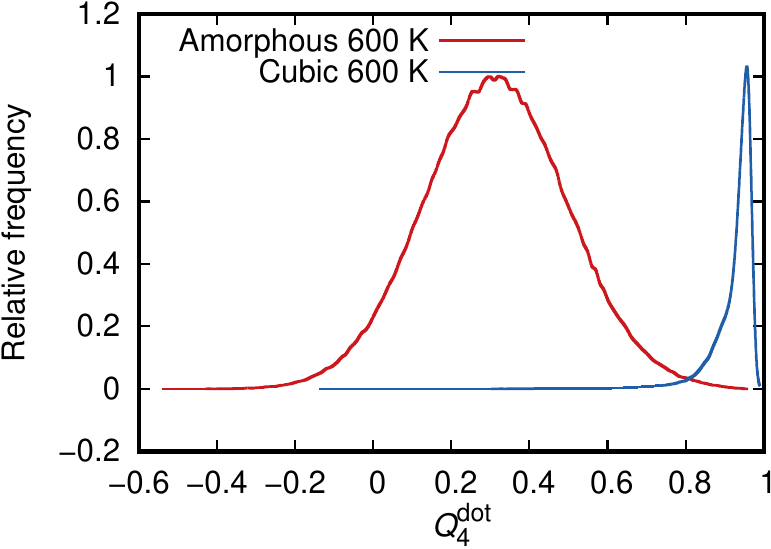}
  \caption{Distribution of the $Q_4^{\rm dot}$ order parameter  for   the liquid and the cubic phase of GST225 at 600 K.
  The order parameter for atom $i$ is defined by \onlinecite{q4}   $Q_{4}^{dot}(i)=\frac{1}{N_i}\sqrt{\sum_{j=1}^{N_i} \sum_{m=-4}^{4}q_{4m,i} q{^*}_{4m,j}}$ where $j$ runs over the $N_i$ neighbors up to a given cutoff (3.6 \AA\ in our case) and  $q_{4m,i}=\frac{1}{N_i}\sum_{j=1}^{N_i}Y_{4m}\left(\mathbf{r}_{ij}\right)$ where $Y_{4m}$ is the fourth order spherical harmonic with order $m$, $ \mathbf{r}_{ij}$ is the vector connecting the two atoms. We chose a threshold of $Q^{dot}_4$=0.87 to identify  a crystalline  atom. 
} 
  \label{fig:Q4}
\end{figure}

\begin{center}

\begin{figure}[h]
 \centering
 
 \includegraphics[width=0.32\textwidth, keepaspectratio]{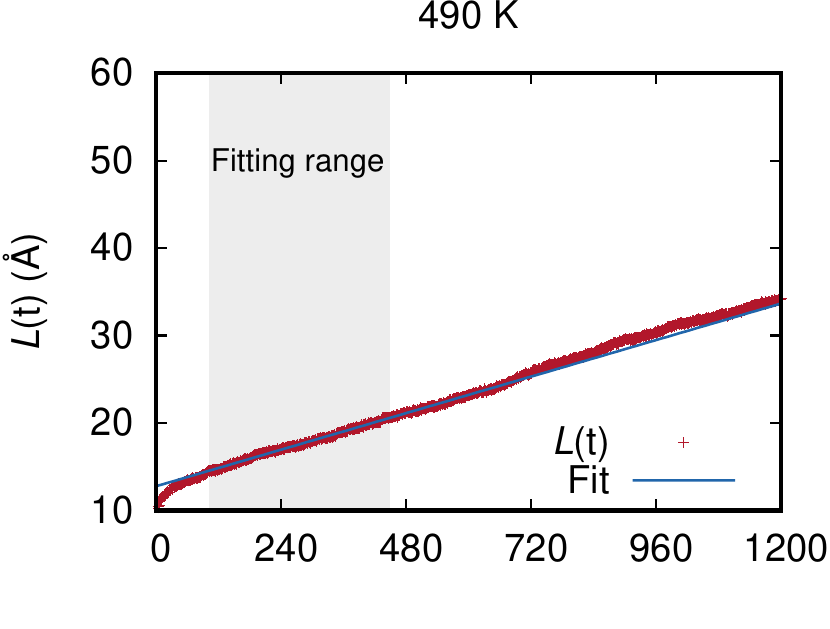}
 \includegraphics[width=0.32\textwidth, keepaspectratio]{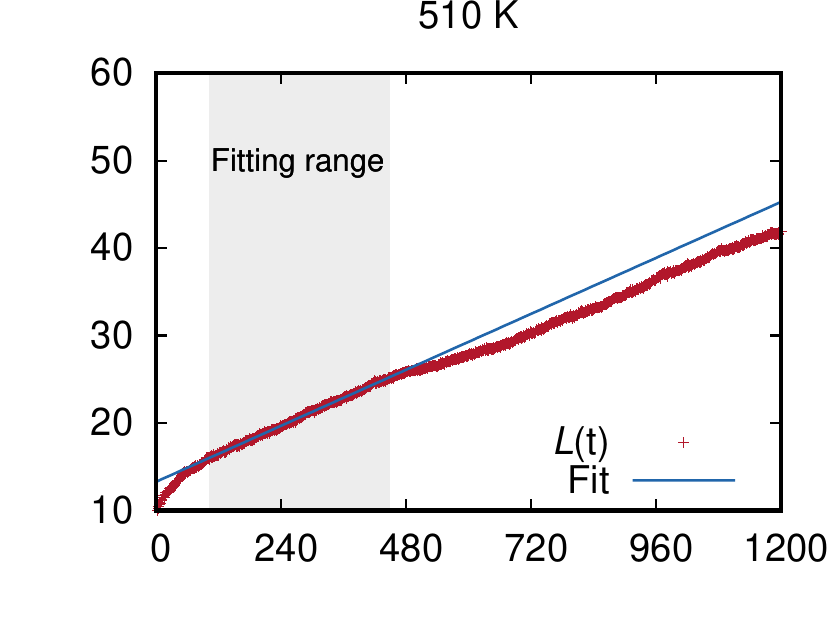}
 \includegraphics[width=0.32\textwidth, keepaspectratio]{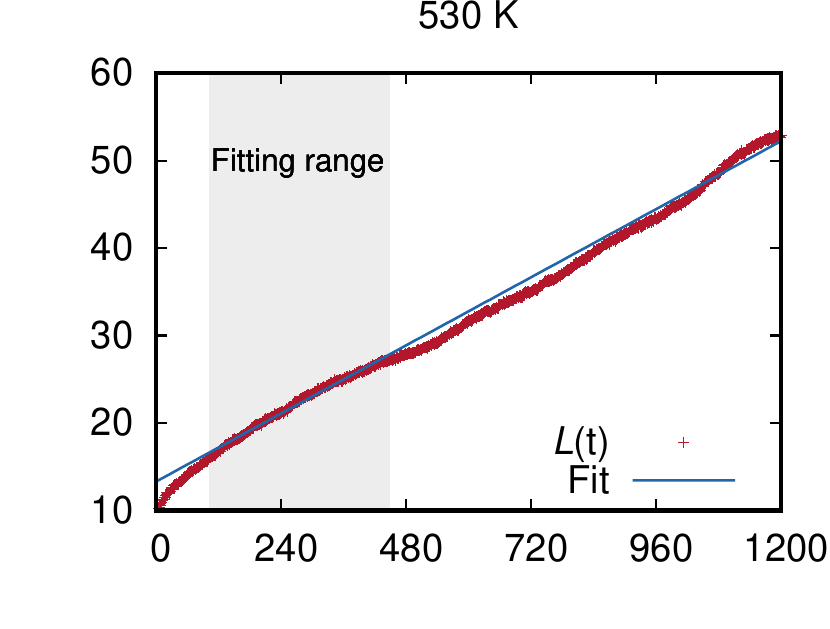}
 \includegraphics[width=0.32\textwidth, keepaspectratio]{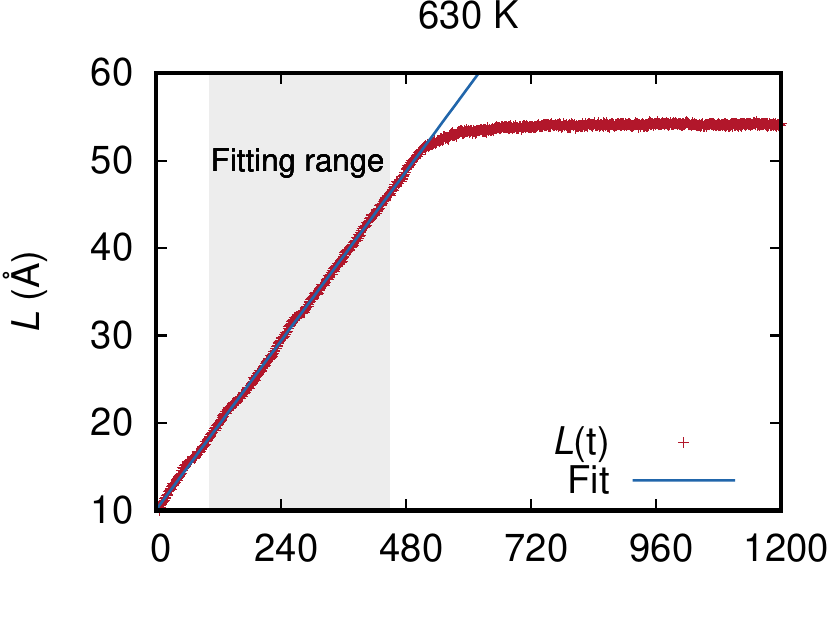}
 \includegraphics[width=0.32\textwidth, keepaspectratio]{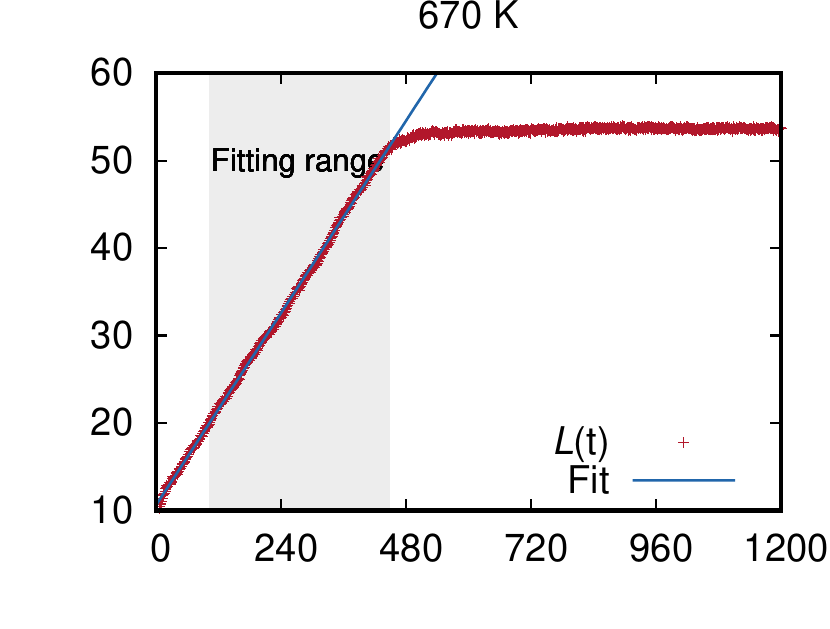}
 \includegraphics[width=0.32\textwidth, keepaspectratio]{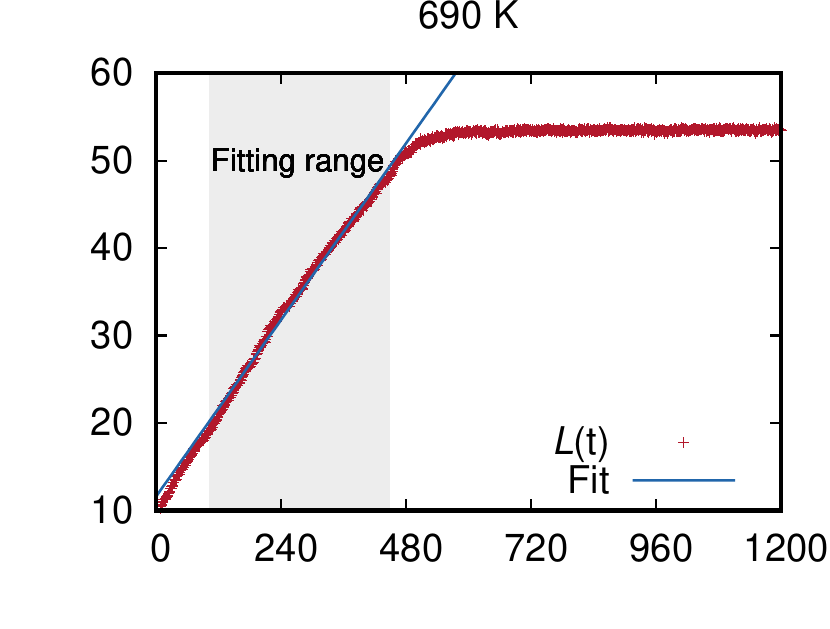}
 \includegraphics[width=0.32\textwidth, keepaspectratio]{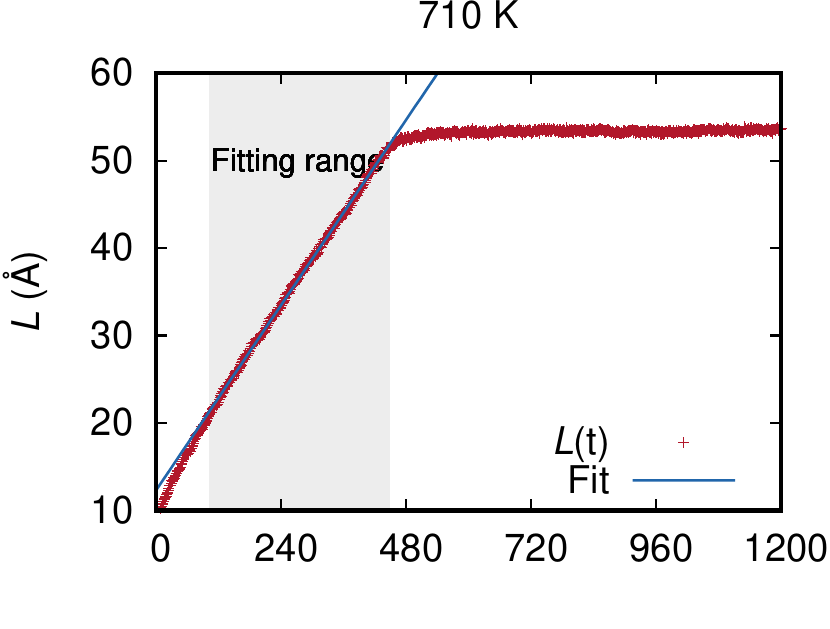}
 \includegraphics[width=0.32\textwidth, keepaspectratio]{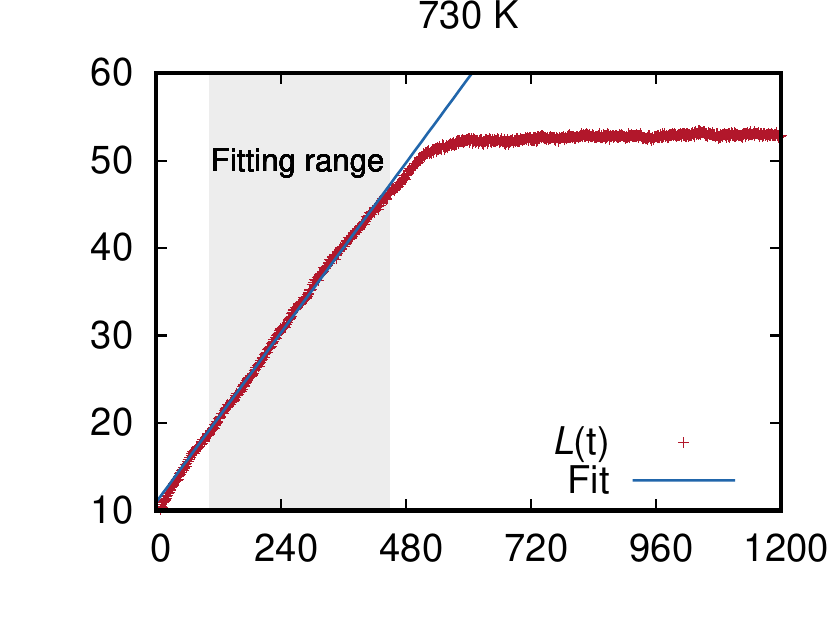}
 \includegraphics[width=0.32\textwidth, keepaspectratio]{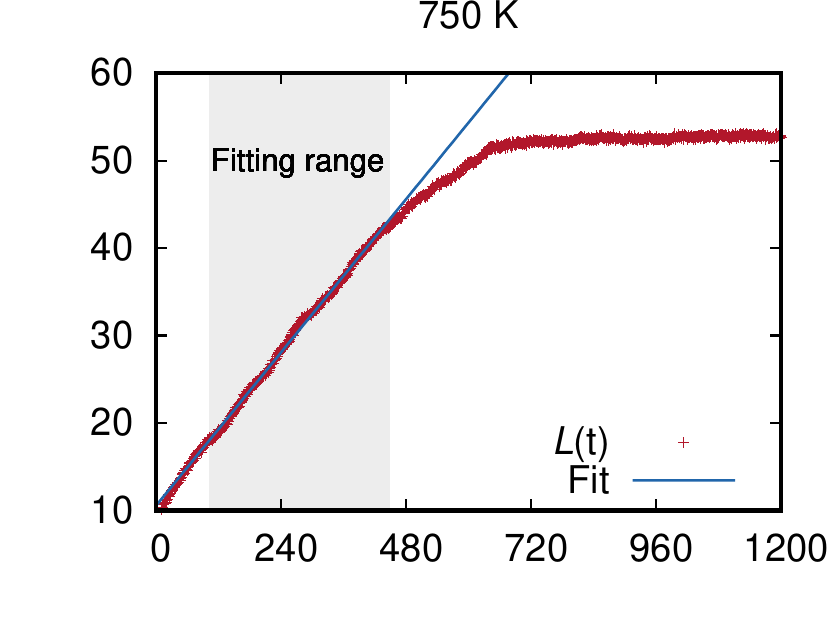}
 \includegraphics[width=0.32\textwidth, keepaspectratio]{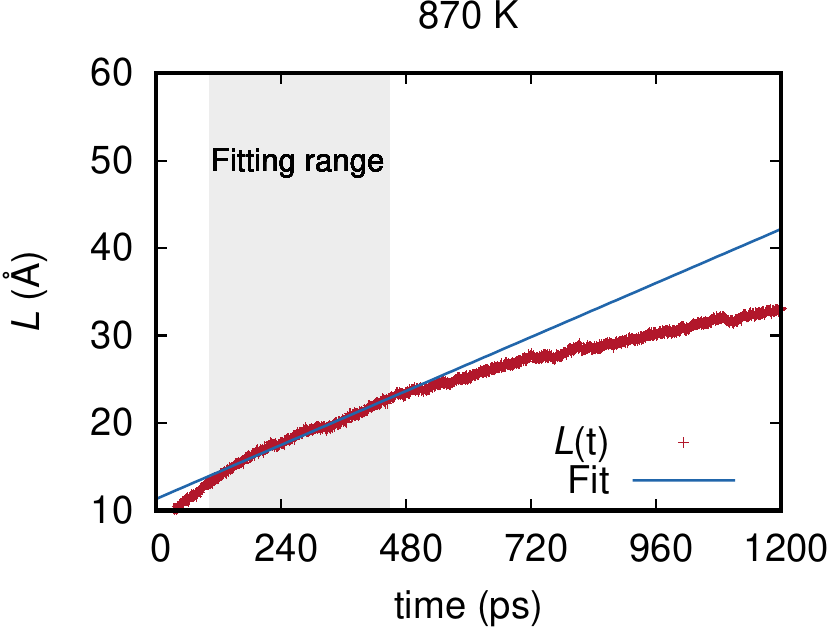}
 \includegraphics[width=0.32\textwidth, keepaspectratio]{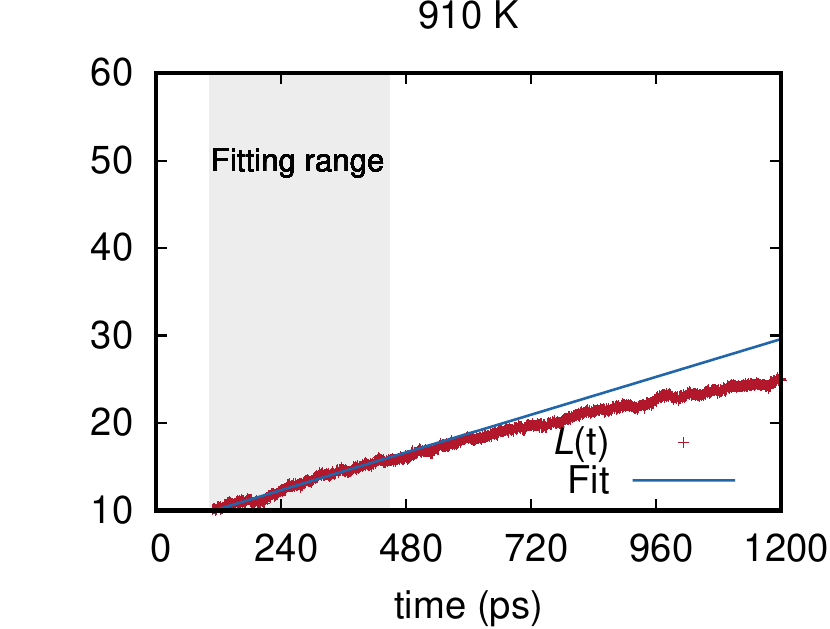}
 \includegraphics[width=0.32\textwidth, keepaspectratio]{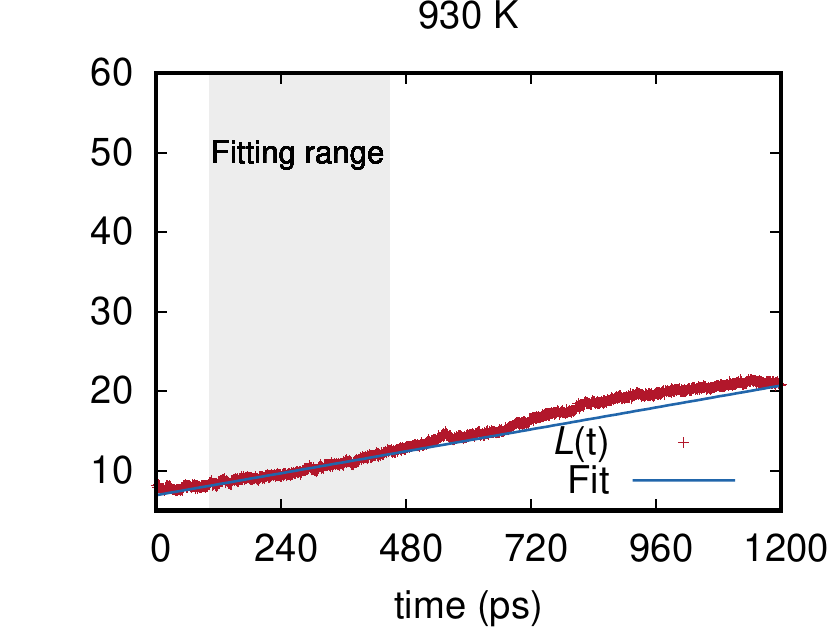}
\caption{The evolution of the 
 the effective (half) 
thickness of the crystalline slab $L(t)$ in simulations at different temperatures and their linear fit in the range highlighted in grey which yields the crystal growth velocity. $L(t)$  is given by $L\left(t\right)=N(t)/({2A\rho_{\rm NN}^{\rm cubic}})$, where $N(t)$ is the number of atoms in the crystalline slab, $\rho_{\rm NN}^{\rm cubic}$ is the theoretical equilibrium density of the cubic phase, $A$ is the cross-section of the cell orthogonal to the growth direction, and the factor two at the denominator accounts for the presence of two growing surfaces.}
 \label{fig:EVO}
\end{figure}
\end{center}{}

\begin{figure}[h]
 \centering
 
\includegraphics[width=0.6\textwidth,keepaspectratio]{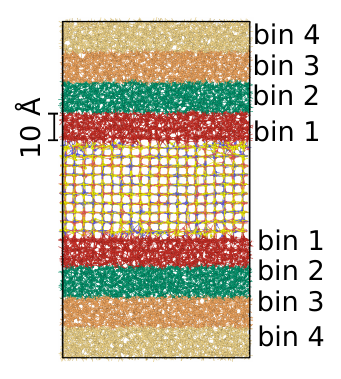}
\caption{ Snapshot of the slices (bins) 10 \AA\ wide at different distances from the interface with the crystalline slab over which we computed the local temperature in \autoref{localT1}-\protect\ref{localT5} and the local diffusion coefficient in \autoref{localD}.}
\label{slices}
\end{figure}

\begin{figure}[h!]
 \centering
 
 \includegraphics[width=0.8\textwidth,keepaspectratio]{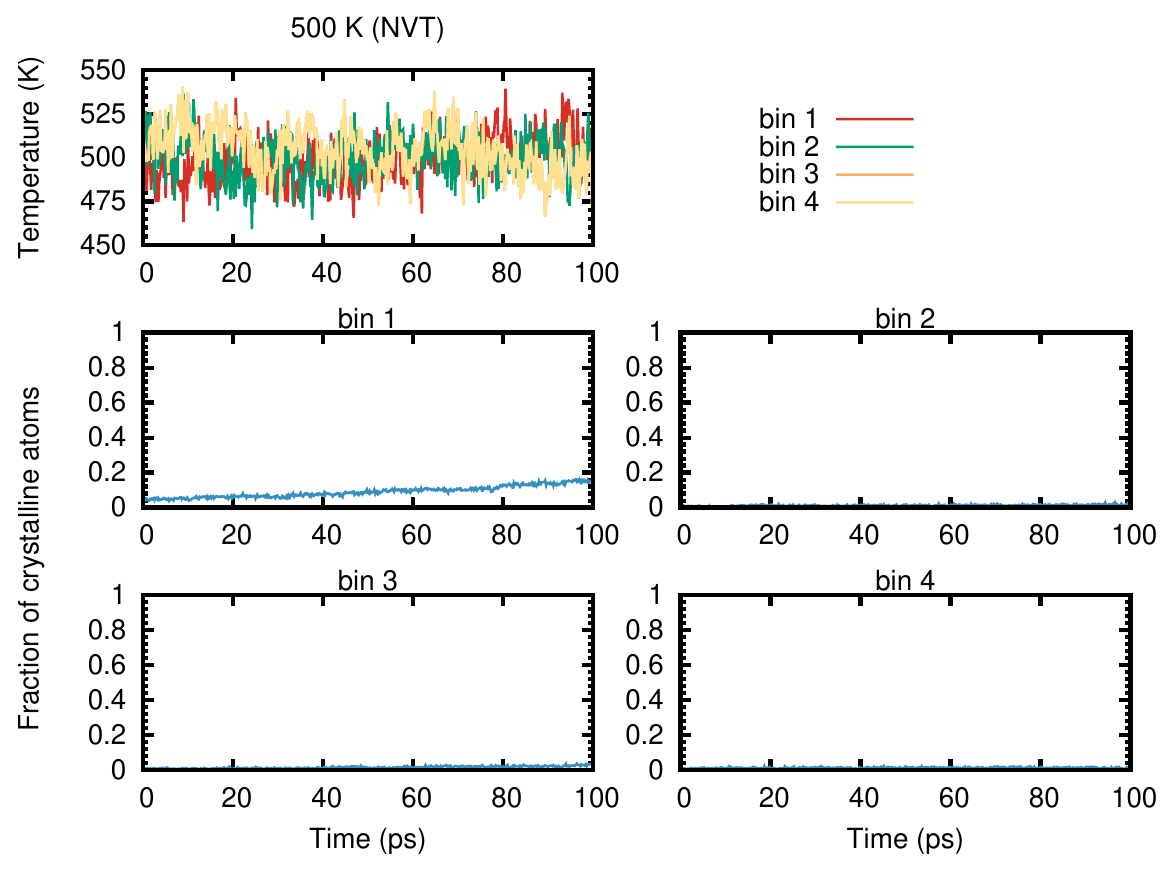}
\caption{ Local temperature and fraction of crystalline atoms as a function of time  in the slices (bins) of \autoref{slices} at different distances from the crystal-liquid interface. 
The data refers to simulations at the global average temperature of 500 K.
Bin 1 is the closest to the interface.}
\label{localT1}
\end{figure}
\clearpage

\begin{figure}[h!]
 \centering
 
 \includegraphics[width=0.8\textwidth,keepaspectratio]{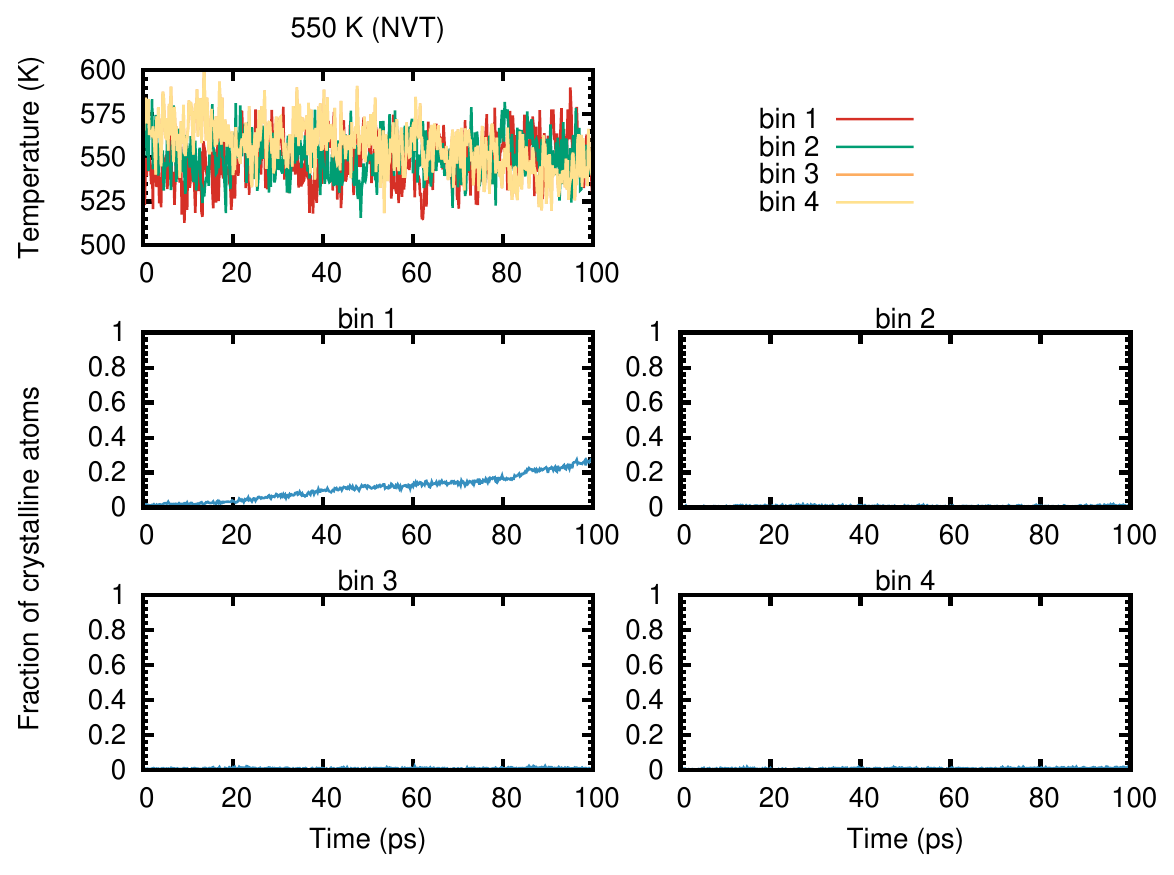}
\caption{Local temperature and fraction of crystalline atoms as a function of time  in the slices (bins) of \autoref{slices} at different distances from the crystal-liquid interface. 
The data refers to simulations at the global average temperature of 550 K.
Bin 1 is the closest to the interface.}
\label{localT2}
\end{figure}

\begin{figure}[h]
 \centering
 
 \includegraphics[width=0.8\textwidth,keepaspectratio]{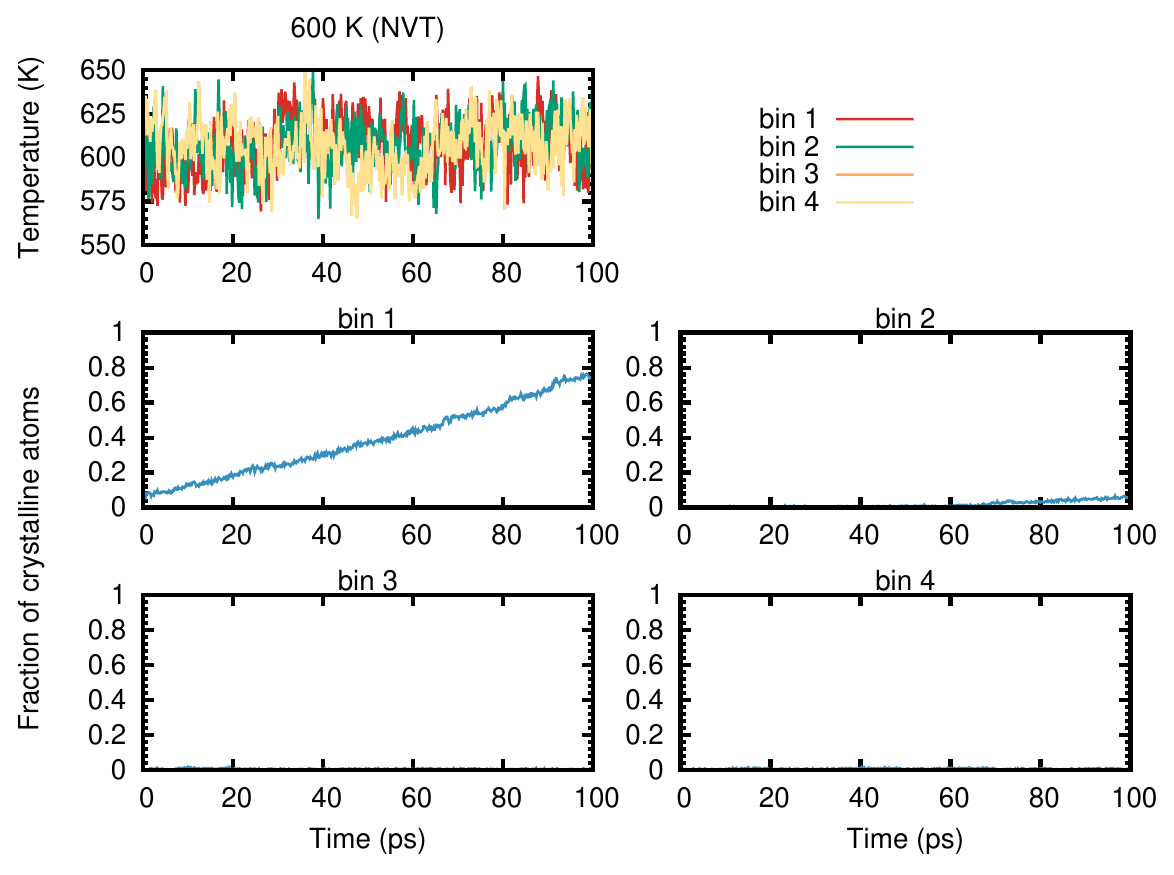}
\caption{Local temperature and fraction of crystalline atoms as a function of time  in the slices (bins) of \autoref{slices} at different distances from the crystal-liquid interface. 
The data refers to simulations at the global average temperature of 600 K.
Bin 1 is the closest to the interface.}
\label{localT3}
\end{figure}

\begin{figure}[h]
 \centering
 
 \includegraphics[width=0.8\textwidth,keepaspectratio]{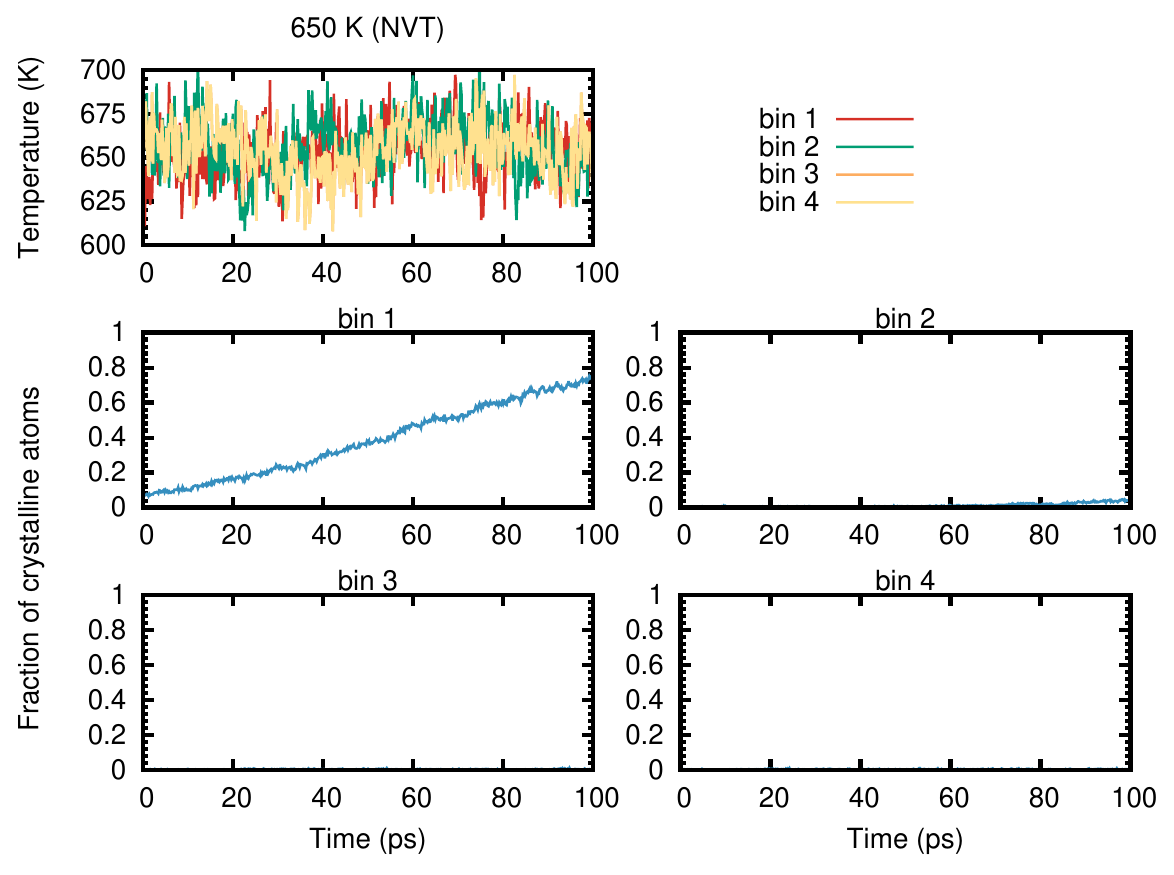}
\caption{ Local temperature and fraction of crystalline atoms as a function of time  in the slices (bins) of \autoref{slices} at different distances from the crystal-liquid interface. 
The data refers to simulations at the global average temperature of 650 K.
Bin 1 is the closest to the interface.}
\label{localT4}
\end{figure}

\begin{figure}[h!]
 \centering
 
 \includegraphics[width=0.8\textwidth,keepaspectratio]{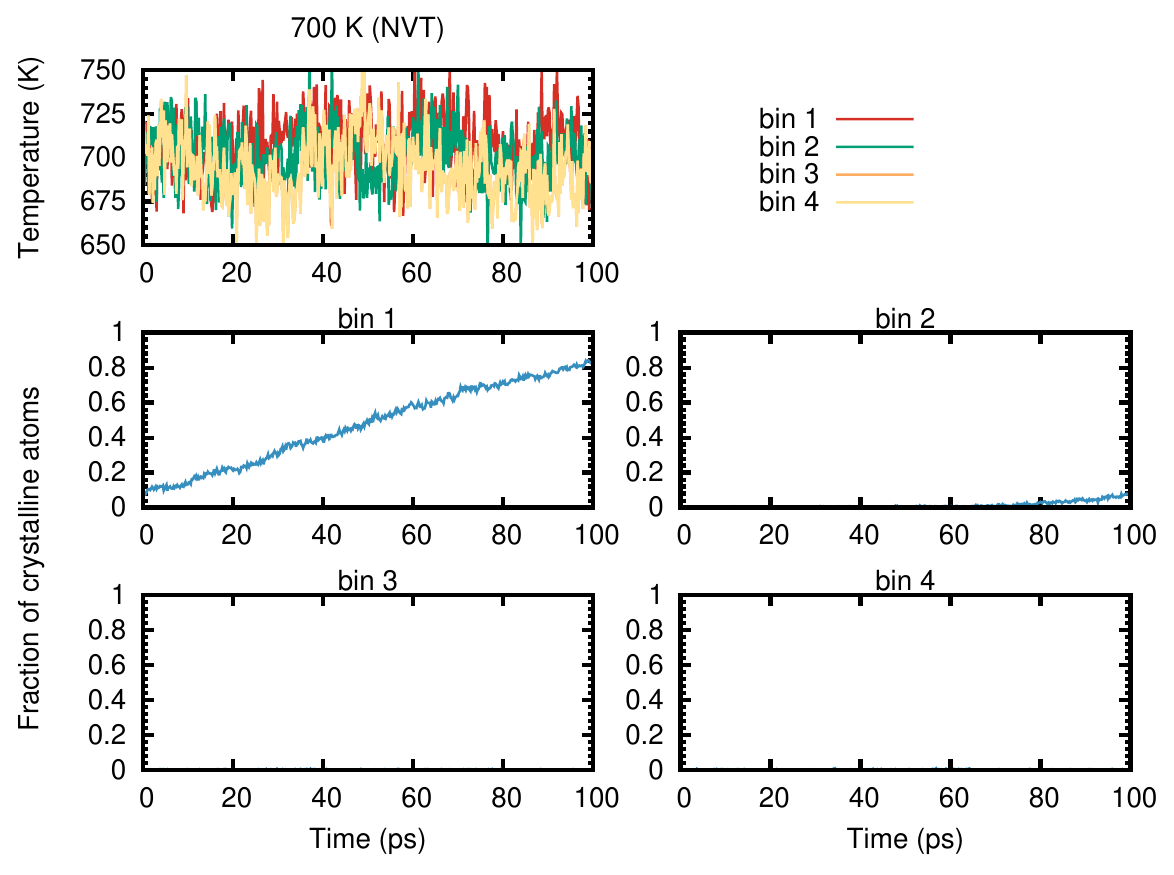}
\caption{ Local temperature and fraction of crystalline atoms as a function of time  in the slices (bins) of \autoref{slices} at different distances from the crystal-liquid interface. 
The data refers to simulations at the global average temperature of 700 K.
Bin 1 is the closest to the interface.}
\label{localT5}     
\end{figure}

\clearpage

\begin{figure}[h!]
 \centering
 
\includegraphics[width=0.8\textwidth,keepaspectratio]{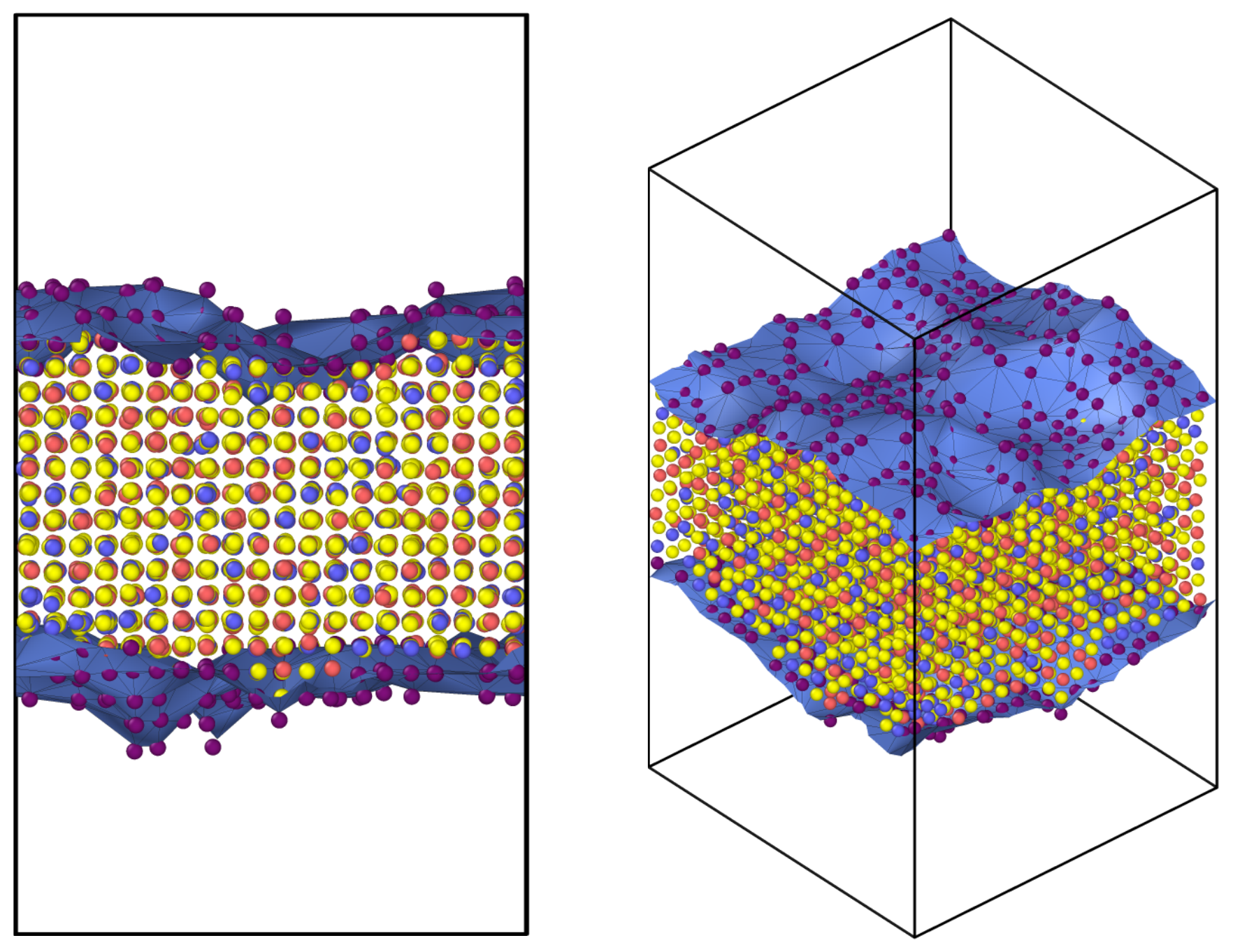}
\caption{ A snapshot of the (100) growing surface at 650 K from two different views. The surface contour is computed with the code Ovito \cite{Ovito}. Only crystalline atoms are shown. Atoms on the liquid-crystal interface are depicted by violet. }
\label{growingsurface}
\end{figure}
\clearpage

\begin{figure}
    \centering
    
\includegraphics[width=0.8\textwidth,keepaspectratio]{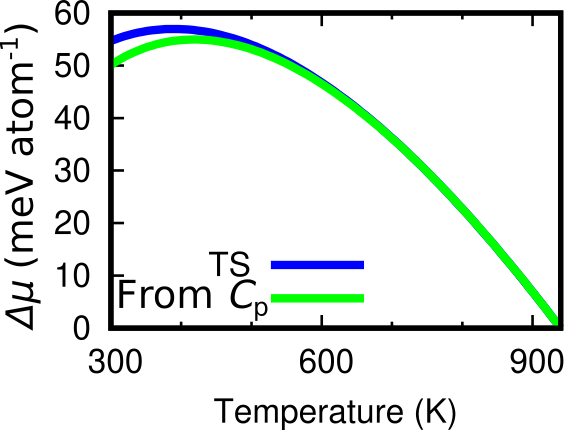}
    \caption{The free energy difference $\Delta\mu$ between the cubic crystal and the liquid as a function of temperature from the Thompson-Spaepen formula (TS) and from the integration of the heat capacity $C_{\rm p}$ (see article).}
    \label{deltamu}
\end{figure}
\clearpage

\begin{figure}[h]
    \centering
    
\includegraphics[width=\textwidth,keepaspectratio]{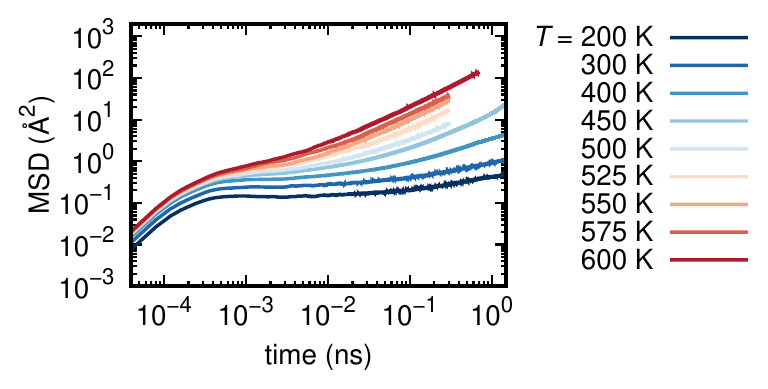}
 \caption{ Mean square displacement (MSD) of GST225 as a function of time at different temperatures  from NVE simulations in the supercooled liquid. At and below 400 K the system is in the amorphous phase. 
For the sake of comparison we mention that the MSD for the metastable cubic phase at 600 K shows a plateau at 0.206 \AA$^2$\ which is slightly higher than the  value obtained in the harmonic approximation from DFT calculation in Ref. \onlinecite{caravati2009first} (Fig. 21 therein).}
    \label{MSD}
\end{figure}

\newpage
\begin{center}
\begin{figure}[h]
    \centering
    \includegraphics[width=0.85\textwidth,keepaspectratio]{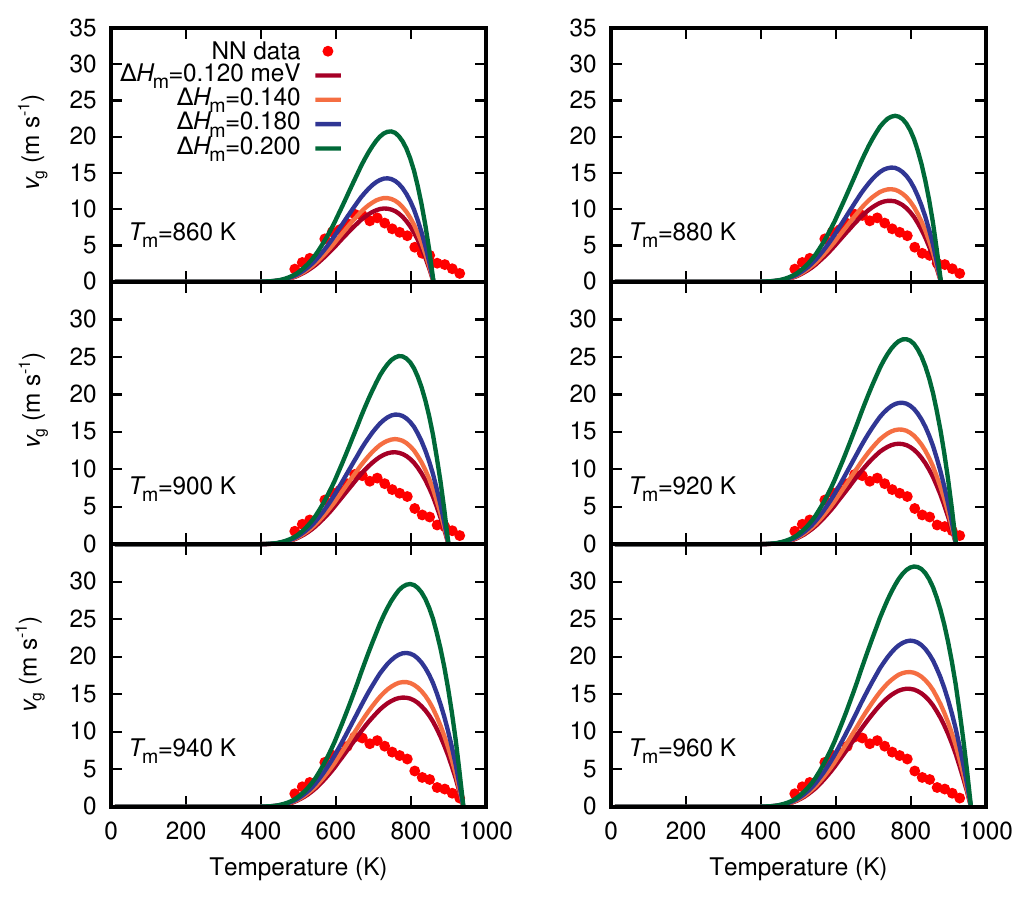}

    \caption{Crystal growth velocity fitted with WF formula and the Thompson-Spaepen expression for $\Delta\mu$ with $T_{\rm m}$ and $\Delta H_{\rm m}$ in the range 860-960 K and 120-200 meV atom$^{-1}$. Each panel corresponds to a different choice of $T_{\rm m}$ and shows the WF fit with different values of $\Delta H_{\rm m}$. The diffusion coefficient $D$ is extracted from the MD simulations and fitted by the Cohen-Grest formula (see Supplementary Figure 1, $\lambda$ is set to 2.42 \AA, see article).}
    \label{fig:my_label}
\end{figure}
\end{center}

\begin{figure}[h]
    \centering
    
     \includegraphics[width=0.6\textwidth,keepaspectratio]{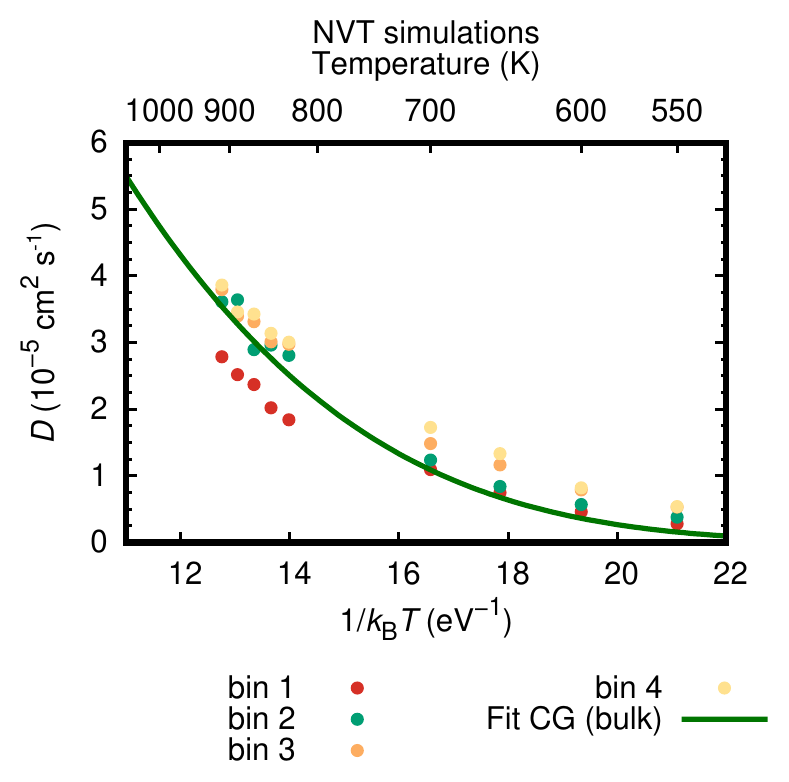}

    \caption{ Local 2D diffusion coefficient in the slices (bins) at different distances from the crystal-liquid interface shown in \autoref{slices}. The local $D$ is computed from the Einstein relation $<x^2>+<y^2>$= 4$Dt$ where $x$ and $y$ lay in the interface plane. The local $D$ is evaluated in runs 20 ps long by averaging over atoms that remain in the same slice over this time span. Bin 1 is the closest to the interface. Data are reported at different temperatures from NVT simulations with the same thermostat used in the study of the crystallization process. The green line corresponds to the bulk values from Supplementary Figure 1.  At low temperatures, the values of $D$ in the bins are lower than the value in the bulk (green line) because they are computed on a shorter time scale due to the fast crystallization from the crystal/liquid interface. In the bulk, $D$ can be computed on a longer time scale because of the presence of an incubation time for the formation of crystalline nuclei.}
    \label{localD}
\end{figure}
\newpage
\begin{center}

\begin{figure}[h]
 \centering
    \includegraphics[width=0.85\textwidth,keepaspectratio]{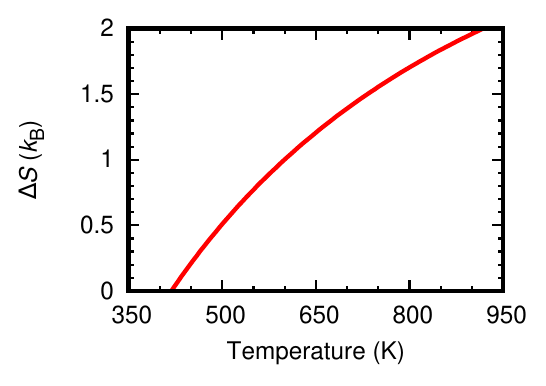}
 \caption{ The difference in entropy between the liquid and the crystal $\Delta S$ as a function of temperature obtained as $\Delta S= \Delta H_{\rm m}/T_{\rm m} - \int^{T_{\rm m}}_T\frac{\Delta C_{\rm v}}{T}dT$, with $\Delta C_{\rm v}$ given by the caloric curves in Fig. 3c in the article.}
 \end{figure}
\end{center}

\clearpage

\begin{figure}[h]
 \centering
     \includegraphics[width=0.85\textwidth,keepaspectratio]{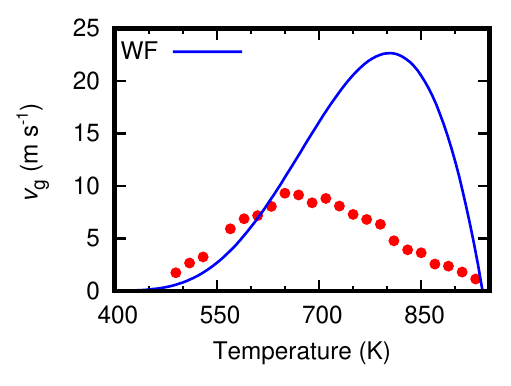}
\caption{ Crystal growth velocities ($v_{\rm g}$) extracted from the motion of the crystal-liquid interface in the MD simulations along the [100] direction of the cubic phase (red dots), fitted with the WF-like formula with the additional factor $exp(\Delta \mu/2k_{\rm B}T$) proposed in Ref. \cite{kelton_greer} and $\lambda$= 2.02 \AA.}  
\end{figure}

\clearpage

\begin{center}

\begin{figure}[h]
 \centering
\includegraphics[width=0.8\textwidth, keepaspectratio]{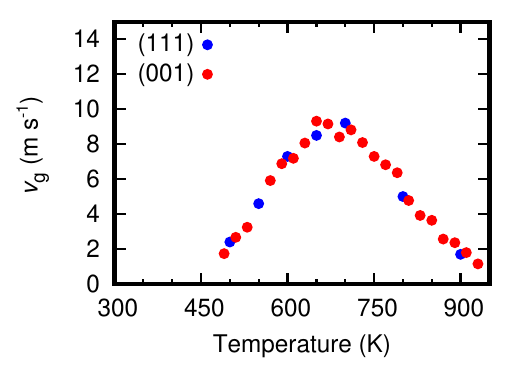}
\caption{ Crystal growth velocities ($v_{\rm g}$) extracted from the motion of the crystal-liquid interface in the MD simulations along the [111] direction of the cubic phase (blue dots). The values of ($v_{\rm g}$) for the (100) surface (see Fig. 6a in the article) are reported for the sake of comparison (red dots).
The simulations for the (111) surface are performed with a 13689-atom supercell and a cross-section of A=2809 \AA$^2$ orthogonal to the growth direction.}  
\end{figure}
\end{center}

\clearpage

\begin{center}

\begin{figure}[h]
 \centering
 
 \includegraphics[width=0.32\textwidth, keepaspectratio]{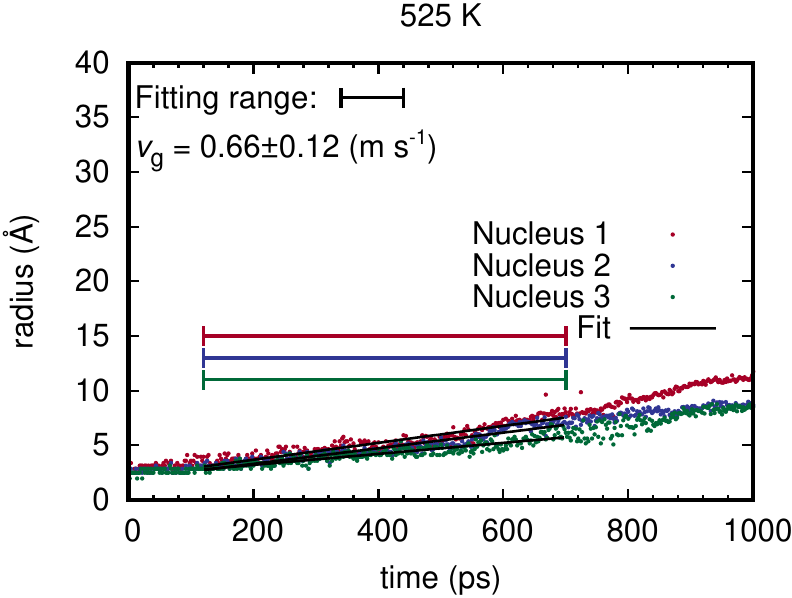}
 \includegraphics[width=0.32\textwidth, keepaspectratio]{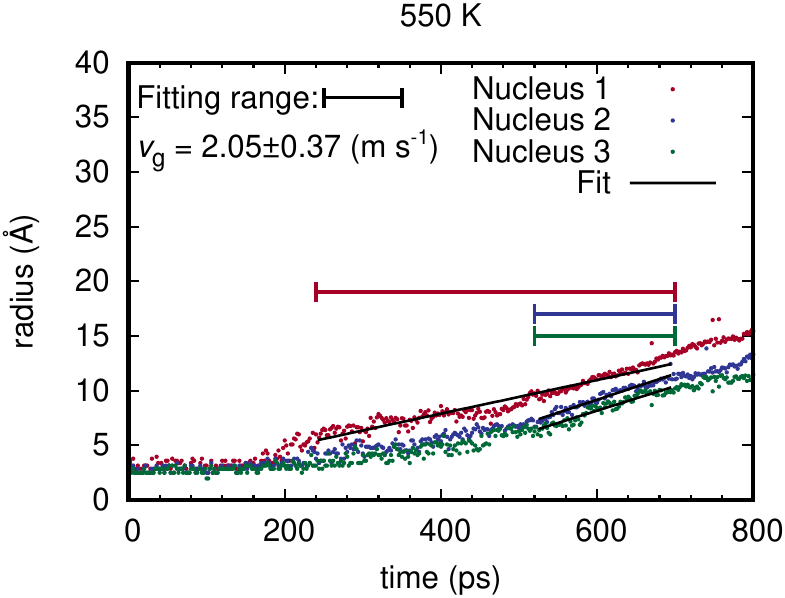}
 \includegraphics[width=0.32\textwidth, keepaspectratio]{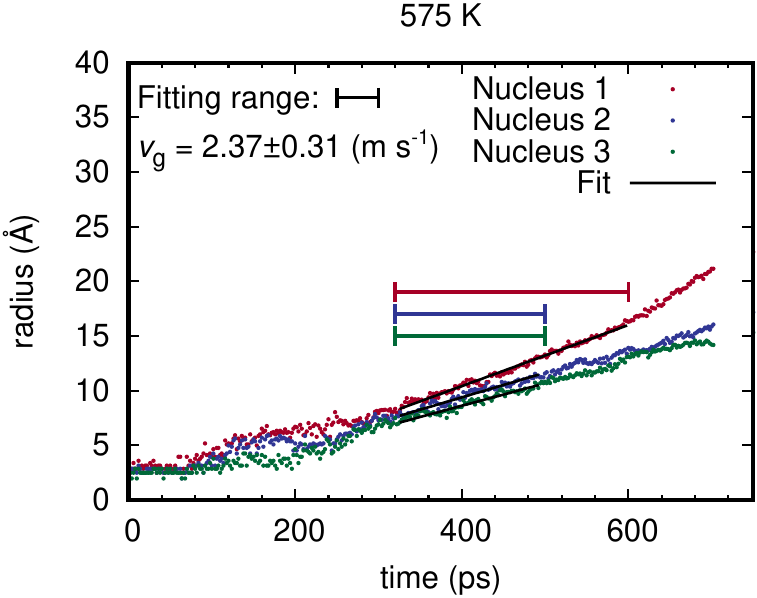}
 \includegraphics[width=0.32\textwidth, keepaspectratio]{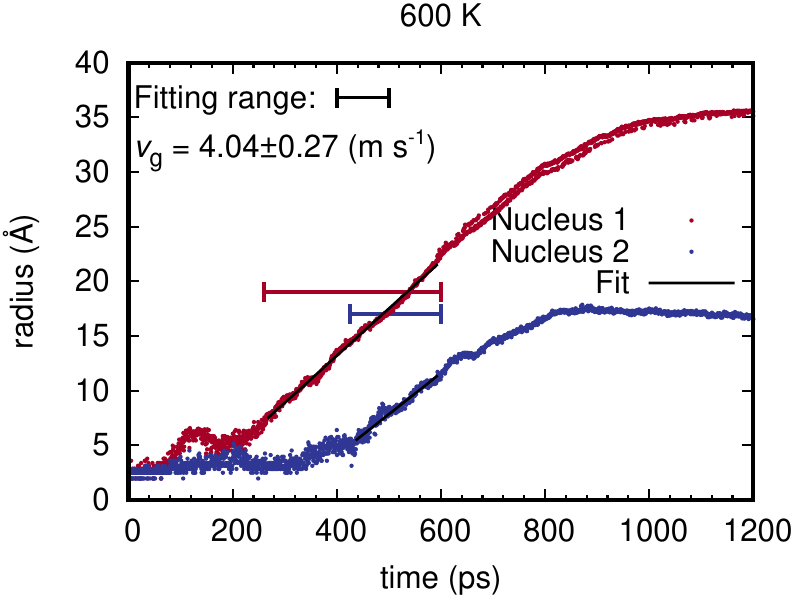}
 \includegraphics[width=0.32\textwidth, keepaspectratio]{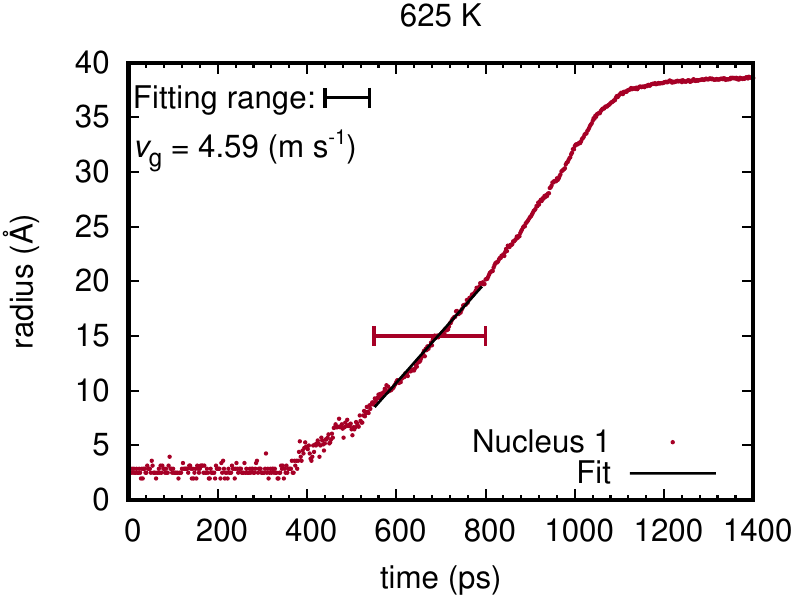}
 \includegraphics[width=0.32\textwidth, keepaspectratio]{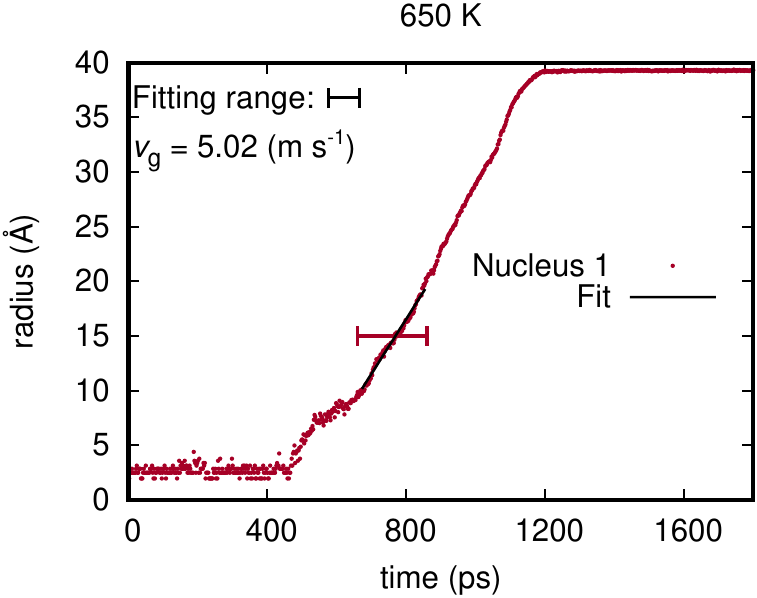}
  \includegraphics[width=0.32\textwidth, keepaspectratio]{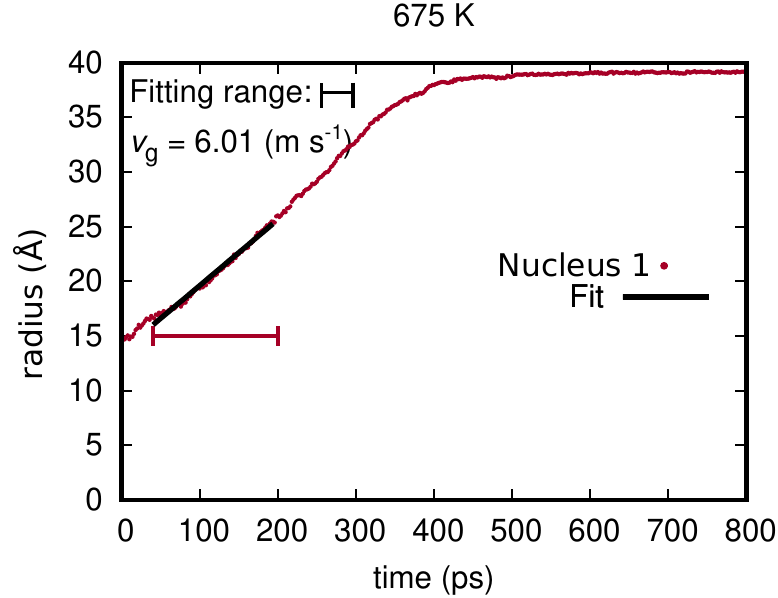}
 \includegraphics[width=0.32\textwidth, keepaspectratio]{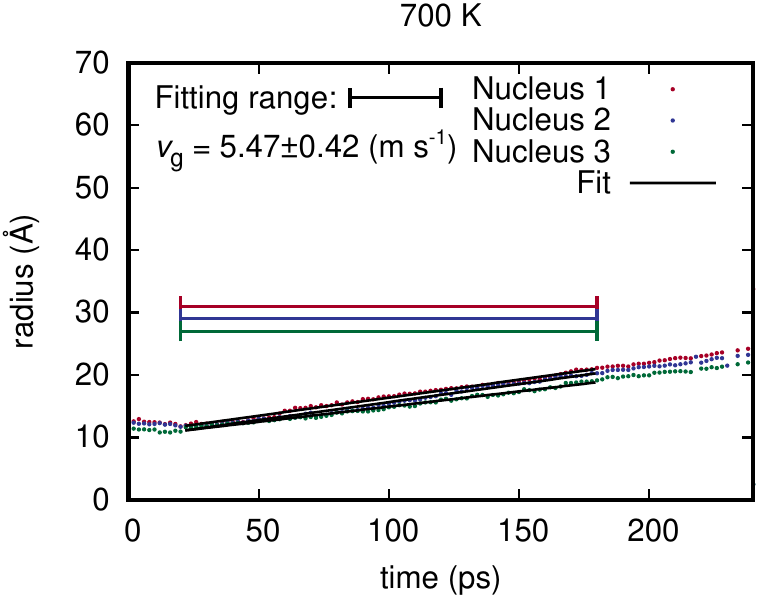}
 \includegraphics[width=0.32\textwidth, keepaspectratio]{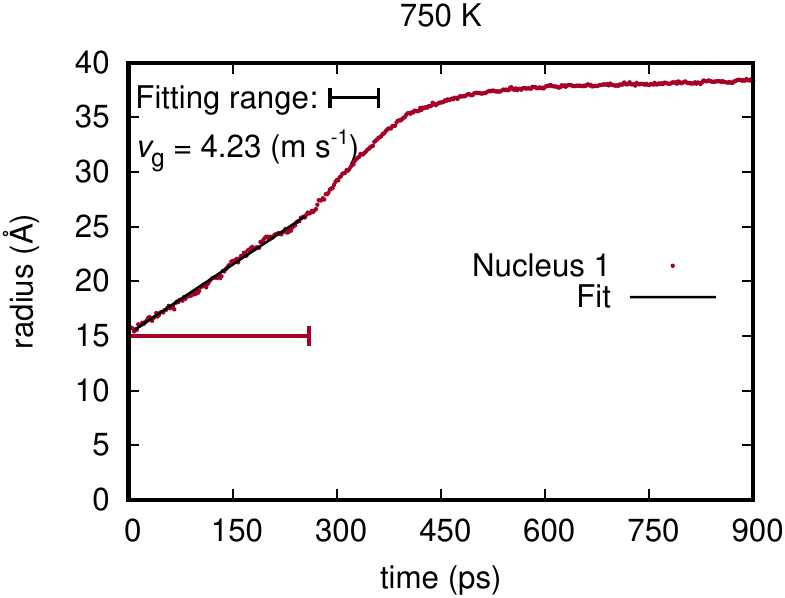}
 \includegraphics[width=0.32\textwidth, keepaspectratio]{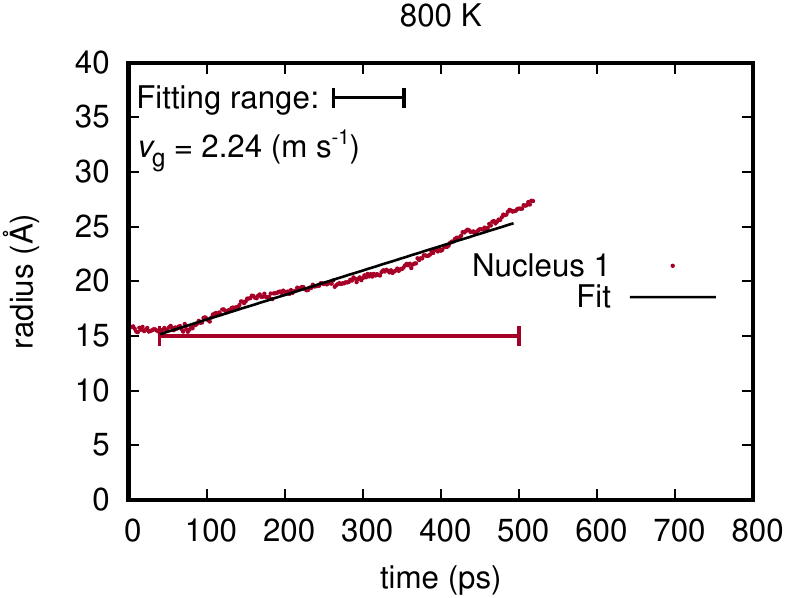}

 \caption{The evolution of the radius  of the crystalline nuclei  as a function of time in simulations at different temperatures in the supercooled liquid and their linear fit in the range highlighted by horizontal bars which yields the crystal growth velocity. In the simulations at and above 675 K no crystal nucleation is observed. The data at these temperatures refer to an overcritical nucleus formed at a lower temperature which was then brought to the higher target temperature to monitor its growth.}
\end{figure}
\end{center}{}
\clearpage

\begin{figure}[h]
 \centering
 
\includegraphics[width=0.8\textwidth, keepaspectratio]{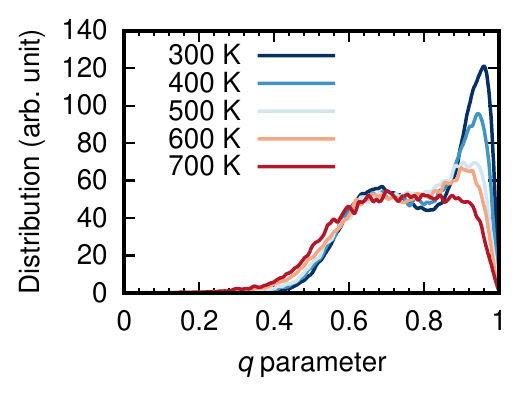}
\caption{Distribution of the local order parameter $q$ for  tetrahedricity  (see article) for four-coordinated Ge atoms in the amorphous phase at different temperature below and above $T_{\rm g}$. The data are obtained by averaging over a time span always shorter than the incubation time for crystallization.}
\label{qparamliquid}
\end{figure}

\begin{center}

\begin{figure}[h!]
 \centering
 
 \includegraphics[width=0.32\textwidth, keepaspectratio]{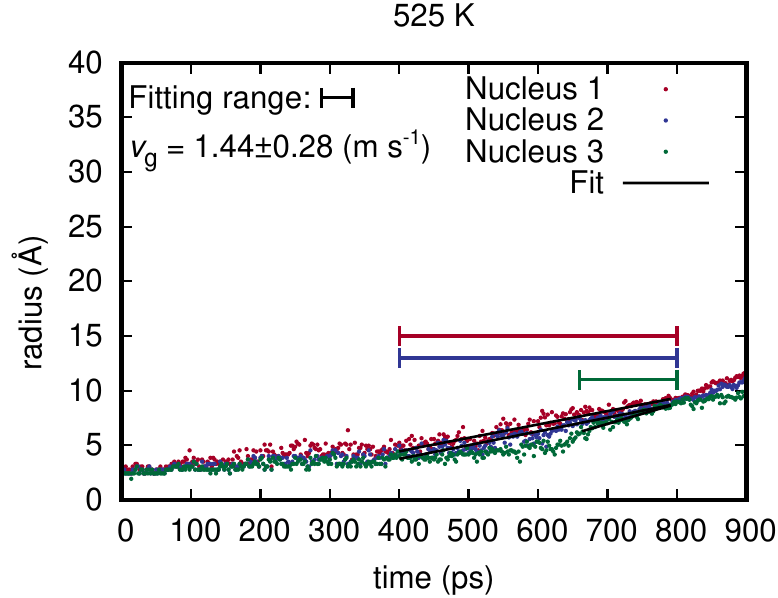}
 \includegraphics[width=0.32\textwidth, keepaspectratio]{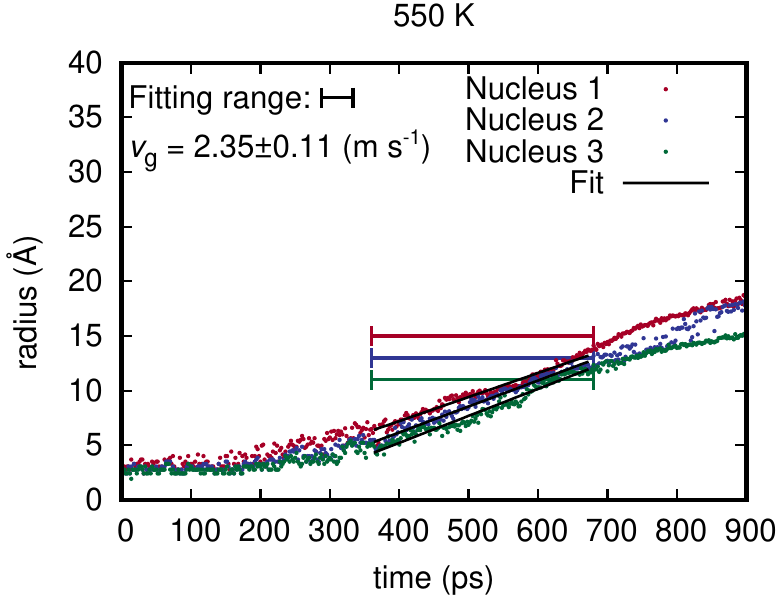}
 \includegraphics[width=0.32\textwidth, keepaspectratio]{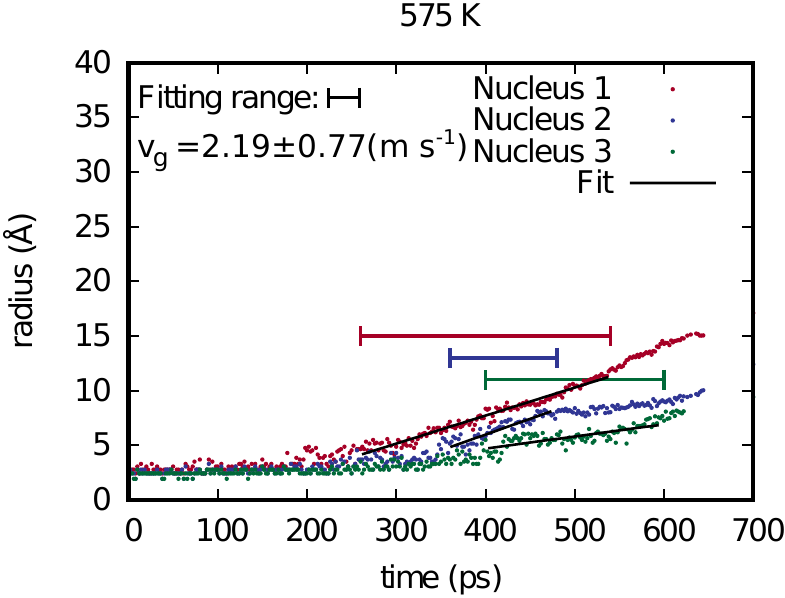}
 \includegraphics[width=0.32\textwidth, keepaspectratio]{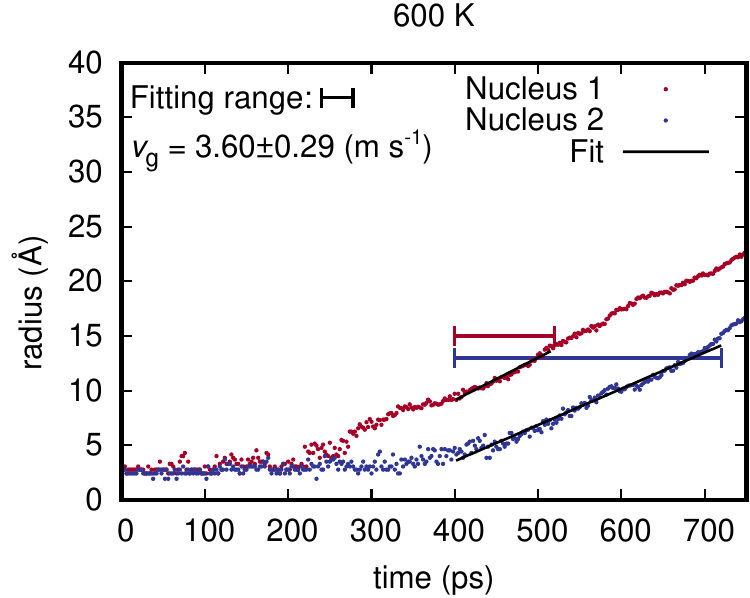}
 \includegraphics[width=0.32\textwidth, keepaspectratio]{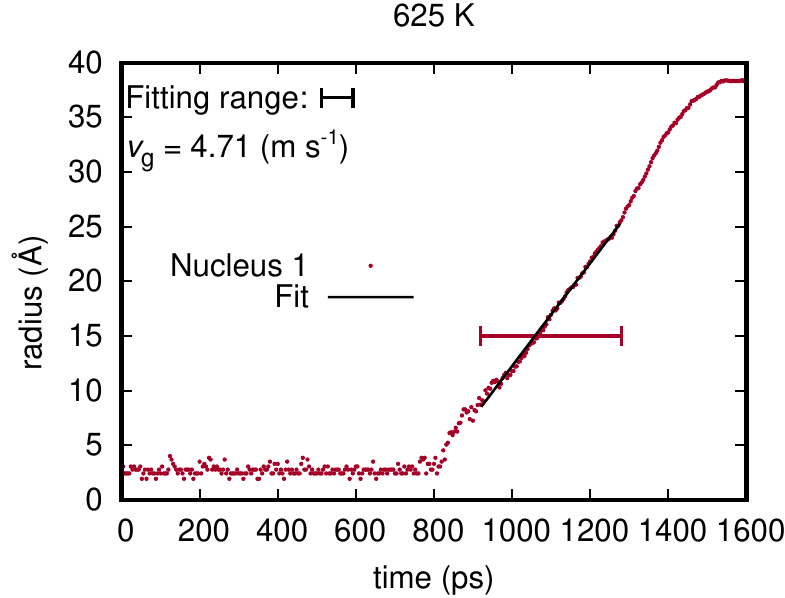}
 \includegraphics[width=0.32\textwidth, keepaspectratio]{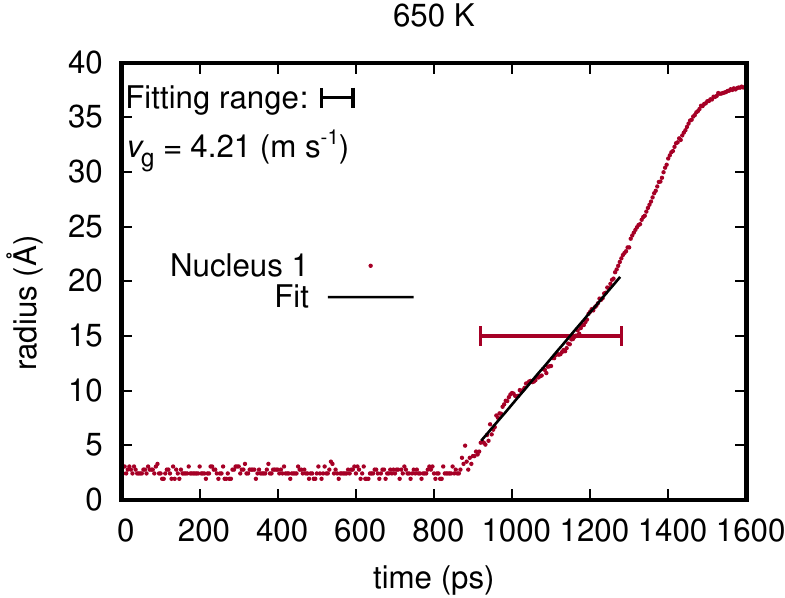}

 \caption{The evolution of the radius  of the crystalline nuclei  as a function of time in simulations at different temperatures in the overheated amorphous phase and their linear fit in the range highlighted by horizontal bars yielding the crystal growth velocity.}
\end{figure}
\end{center}{}

\newpage
\begin{table}[h!]
\centering
\captionof{table}{Details of the hyperparameters
used for the learning rate and the Loss Function (LF) employed for the training of the NN potential as defined in Ref. \cite{wang2018deepmd}.}
\label{tab:paramtri}
\begin{tabular}{l l l }

\hline
\hline

Learning rate & initial value  = 0.001 & final value =   3.51 10$^{-8}$ \\
Decay of the learning rate& prefactor = 0.95  & steps = 7275     \\
Energy prefactor in the LF& initial value = 1  & fin    al value = 1000 \\
Forces prefactor in the LF& initial value = 8000  & final value = 800  \\
Virial prefactor in the LF& initial value = 0.2  & final value = 100  \\
Batch size (configurations)& training = 16   & test = 128    \\
Training epochs& 200  &   \\

\hline
\hline
\end{tabular}
\end{table}

\bibliography{ref2.bib}